\documentclass{article}
\usepackage{tgpagella}
\usepackage{mathpazo}
\usepackage[utf8]{inputenc}
\usepackage{amsmath}
\usepackage{amsthm}
\usepackage{mathtools}
\usepackage{amssymb}
\usepackage{nicefrac}
\usepackage{xcolor}
\usepackage[hidelinks]{hyperref}
\usepackage{tikz}
\usepackage{enumerate}
\usepackage[top=1in,bottom=1in,right=1in,left=1in]{geometry}
\usepackage{physics}
\usepackage{bbm}
\usepackage[symbol]{footmisc}
\usepackage{epigraph}
\usepackage[english]{babel}
\usepackage[style=english]{csquotes}
\usepackage{cleveref}
\usepackage[style=philosophy-classic, sorting=nymdt]{biblatex}
\DeclareFieldFormat{pages}{pp. #1.}
\DefineBibliographyStrings{english}{%
  page             = {},
  pages            = {},
} 
\DeclareSortingTemplate{nymdt}{
  \sort{
    \field{presort}
  }
  \sort[final]{
    \field{sortkey}
  }
  \sort{
    \field{sortname}
    \field{author}
    \field{editor}
    \field{translator}
    \field{sorttitle}
    \field{title}
  }
  \sort{
    \field{sortyear}
    \field{year}
  }
  \sort{
    \field[padside=left,padwidth=2,padchar=0]{month}
    \literal{00}
  }
  \sort{
    \field[padside=left,padwidth=2,padchar=0]{day}
    \literal{00}
  }
  \sort{
    \field{sorttitle}
  }
  \sort{
    \field[padside=left,padwidth=4,padchar=0]{volume}
    \literal{0000}
  }
}

\newcommand{\CJ}{\mathcal{J}}

\newcommand{\CS}{\mathcal{S}}
\newcommand{\CE}{\mathcal{E}}
\newcommand{\e}{\mathrm{e}}
\newcommand{\I}{\mathrm{i}}

\newtheorem*{scholium}{Scholium} 

\title{The Dual Dynamical Foundation of Orthodox Quantum Mechanics}
\author{
Diana Taschetto\thanks{Freudental Institute, Department of Mathematics, Faculty of Science, Utrecht University. Contact: d.taschetto@uu.nl} \and Ricardo Correa da Silva\thanks{Department of Mathematics, Friedrich-Alexander-Universität Erlangen-Nürnberg. Contact: ricardo.correa.silva@fau.de}
}
\date{}
\begin{document}
\maketitle
\begin{large}
\begin{center}
     To appear in \textit{Studies in History and Philosophy of Science}  
\end{center}
\end{large}
\begin{abstract}
\noindent This paper combines mathematical, philosophical, and historical analyses in a comprehensive investigation of the dynamical foundations of the formalism of orthodox quantum mechanics. The results obtained include: (i) A deduction of the canonical commutation relations (CCR) from the tenets of Matrix Mechanics; (ii) A discussion of the meaning of Schrödinger's first derivation of the wave equation that not only improves on Joas and Lehner's 2009 investigation on the subject, but also demonstrates that the CCR follow of necessity from Schrödinger's first derivation of the wave equation, thus correcting the common misconception that the CCR were only posited by Schrödinger to pursue equivalence with Matrix Mechanics; (iii) A discussion of the mathematical facts and requirements involved in the equivalence of Matrix and Wave Mechanics that improves on F. A. Muller's classical treatment of the subject; (iv) A proof that the equivalence of Matrix and Wave Mechanics is necessitated by the formal requirements of a dual action functional from which both the dynamical postulates of orthodox quantum mechanics, von Neumann's process 1 and process 2, follow; (v) A critical assessment, based on (iii) and (iv), of von Neumann's construction of unified quantum mechanics over Hilbert space.  Point (iv) is our main result. It brings to the open the important, but hitherto ignored, fact that orthodox quantum mechanics is no exception to the golden rule of physics that the dynamics of a physical theory must follow from the action functional. If orthodox quantum mechanics, based as it is on the assumption of the equivalence of Matrix and Wave Mechanics, has this ``peculiar dual dynamics,'' as von Neumann called it, then this is so because by assuming the equivalence, one has been assuming a peculiar dual action.

\vspace{6pt}
\end{abstract}

\renewcommand{\thefootnote}{[\arabic{footnote}]}

\epigraph{\flushright \textit{Is discontinuity destined to reign over the physical universe, and will its triumph be final? Or will it finally be recognized that this discontinuity is only apparent, and a disguise for a series of continuous processes?}}{H. Poincaré, ``L'hypothèse des Quanta,'' 1913}

\section{Introduction}

This paper is a comprehensive investigation of the dynamical foundations of the formalism of orthodox quantum mechanics. It applies the method of integrating history, mathematical analysis, and philosophy of physics to make maximal sense of foundational questions related to the dynamics of quantum mechanics. Mathematical and conceptual analyses are applied to address concerns relative to the history of quantum mechanics, and historical insights are then used to treat present-day concerns about the foundations of quantum mechanics. The most novel, most essential, of our results is the discovery of the existence of a dual action functional from which both the dynamical postulates of orthodox quantum mechanics necessarily follow. This action functional is proved to be the simple logical condition that necessitates the equivalence of Matrix and Wave Mechanics; we argue, therewith, for the following implication: The dual dynamics of von Neumann's ``unified'' quantum mechanics, process 1 and process 2, does not enter into the formalism of quantum mechanics mysteriously or \textit{ad hoc}: the dual dynamics is completely determined by the dual action.

The investigation is organized as follows. We start by solving a historical problem, namely the conceptual origins of the canonical commutation relations in both Matrix Mechanics (\S2.1) and Wave Mechanics (\S2.2). The solution casts the problem of the equivalence of these theories, first examined by Schrödinger in 1926, in a form that has not been heretofore considered by the literature; we discuss its features in \S2.3. We attack this problem, and find its solution, in \S2.4: The equivalence is logically necessitated by the formal conditions of a dual action functional whose different parts have independently entailed, as a matter of historical and mathematical fact, the emergence of the canonical commutation relations in Matrix and Wave Mechanics. Therewith, consequences of theoretical interest follow, certified by the following fact: The equivalence of Matrix and Wave Mechanics is stated by von Neumann to be \textit{the logical foundation} of his \textit{The Mathematical Foundations of Quantum Mechanics}, and this work is acknowledged by the literature to have consolidated the formalism of orthodox quantum mechanics.

The subsequent discussion in \S3.1 is devoted to a critical analysis of the arguments used by von Neumann to ``unify'' Matrix and Wave Mechanics in Hilbert space. This analysis leads directly to a logical understanding of von Neumann's formulation of the measurement problem that does not conform with the official understanding of the problem in terms of the so-called problem of outcomes \parencite{Maudlin1995}. But then, since the problem  of outcomes has given rise to much dispute in the foundations of quantum mechanics, the cogency of our results will seem to depend on our ability to explain away such discrepancy. We solve this problem in \S3.2. The popular view that von Neumann proposed process 1 only to solve the measurement problem will be refuted, and a new understanding of von Neumann's Impossibility Proofs will thereby emerge, evincing that contemporary charges against von Neumann's dignity as a mathematician, according to which his argumentation is at points ``\textit{ad hoc},'' ``arbitrary,'' and ``silly'' (John Bell), are unfounded.\footnote{\label{foot:Bellimpossibility}{This conclusion is in agreement with the recent literature that reevaluates Bell's classical analysis of von Neumann's Impossibility Proof; see \parencite{bub2010, dieks2017, acuna2021}. }}

 \section{The Dynamical Duality Between Matrix and Wave Mechanics}

 \subsection{The Dynamical Origin of the Quantization Rules, Part I: Matrix Mechanics}

Matrix Mechanics postulates \parencite[292]{Born&Jordan1925} that all the conclusions of the theory rest upon the condition 
\begin{equation} \label{CCR2}
    pq - qp = -\I  \hbar.\quad \footnote{This statement can be readily understood upon recalling the fact that the fundamental problem of Matrix Mechanics is to diagonalize the Hamiltonian matrix, and it is the demand imposed on the diagonalization procedure that the transformation satisfies the CCR (see \parencite[\S3]{Heisenberg&Born&Jordan1925}).}
\end{equation}
Today we refer to it as the ``canonical commutation relation,'' or ``Heisenberg commutation relation;'' the latter attribution, however, is placed in proper historical context by \parencite{Born1955}, who reminisced: ``I shall never forget the thrill I experienced when I succeeded in condensing Heisenberg's ideas on quantum conditions in the mysterious equation $pq - qp = -\I  \hbar$.'' The ``mystery,'' however, seems to ensue from the fact that the premises from which the equation was derived have been omitted by the founding fathers; the ``physical ideas'' of \parencite{Heisenberg1925} are buried in the off-beat mathematical method used in the paper,\footnote{A substantial amount of historical work has been devoted to make this paper intelligible to the modern reader; see, e.g., \parencite{MacKinnon1977} and, for a more recent discussion, \parencite{blum2017}.} and Born himself did not publish a step-by-step calculation that, as he put it, ``condensed'' them.\footnote{The derivation in \parencite[290-291]{Born&Jordan1925} is in itself unintelligible.} Indeed, \eqref{CCR2} was introduced in Matrix Mechanics---and then later, by von Neumann, in orthodox quantum mechanics---as an unexplained postulate; fortunately, however, the obscurity can be eliminated by making the problem attacked in \parencite{Heisenberg1925}, and the premises used to solve it, absolutely explicit. Once we do this, a derivation of \eqref{CCR2}, and therewith a physical understanding of it, will readily follow, as the reader shall see in this section.

To understand the conceptual substratum upon which quantum mechanics was built it is indispensable to remember that older works on the quantum theory\footnote{\cite{Einstein1917}; \cite{Ehrenfest1917}; \cite{Bohr1918}; \cite{Bohr1924}; \cite{Kramers1924a}; \cite{Kramers1924b}; \cite{Born1924}, \cite{KramersandHeisenberg1925}.} prior to \cite{Heisenberg1925} uniformly show the following attitude toward the application of the quantum hypothesis: the goal was to give a mechanical account of microprocesses on the basis of two fundamental assumptions. They are:
\begin{itemize}

\item Assumption I. \textit{Dynamical Discontinuity.} Regardless of classical theory, the quantum-theoretically stable orbits of a Hamiltonian system $H(p_1, ..., p_n; q_1, ..., q_n)$ are characterized by the condition imposed on the reduced action $S_{0}$ as
\begin{equation}\label{Bohr-Sommerfeld}
S_0 = \oint dq_k p_k = hn_k,  
\end{equation}
$k = 1, ..., n$, where $h$ is the Planck constant, and $n_k$ are positive integers labeling the stationary states. Thus the energy of a quantum system \textit{does not run in a continuum}; energy states rather transit in \textit{jumps}, and it is this ``discontinuous motion'' that gives rise to the emission of electromagnetic radiation.\footnote{It is important to remember that, in point of historical fact, photons were only accepted by the scientific community after 1924. It was a fundamental postulate of Bohr's theory of atomic structure that the radiation emitted according to \eqref{Bohrenergycondition} consisted \textit{of a train of simple harmonic waves}; see \parencite[Chapter 1]{Bohr1923}. The incorporation of wave-particle duality into the formalism of the quantum theory is a fascinating, and exceedingly important, aspect of its history that unfortunately had not yet been subject to rigorous and detailed historical investigation; see, however, \parencite{Hendry1980}.}
This relationship between the motion of the system and the emitted radiation is given by the so-called ``Bohr's frequency condition'',
\begin{equation}\label{Bohrenergycondition}
    E(n') - E(n) = h\nu(n', n).
\end{equation}

Thus later, at their talk the 1927 Solvay Conference, M. Born and W. Heisenberg emphasized:
\begin{quote}
    Quantum mechanics is based on the intuition that the essential difference between atomic physics and classical physics is the occurrence of discontinuities. Quantum mechanics should thus be considered a direct continuation of the quantum theory founded by Planck, Einstein, and Bohr. [....] The discontinuous nature of the atomic processes here is put into the theory from the start as empirically given. \parencite[408, 411]{BornandHeisenberg1927}
\end{quote}

Also John von Neumann, in opening his \textit{The Mathematical Foundations of Quantum Mechanics}, was explicit in that the physical ground of his mathematical treatise was taken from the start as a given:
\begin{quote}
    This is not the place to point out the great success which the quantum theory attained in the period from 1900 to 1925, a development which was dominated by the names of Planck, Einstein and Bohr. At the end of this period of development it was clear beyond doubt [!] that \textit{all} elementary processes, i.e., all occurrences of an atomic or molecular order of magnitude, obey the \textit{“discontinuous”} laws of quanta. [...] What was fundamentally of greater significance was that the general opinion in theoretical physics had accepted the idea that \textit{the principle of continuity} (``\textit{natura non facit saltus}''), prevailing in the macroscopic world, \textit{is merely simulated by an averaging process} in a world which \textit{in truth is discontinuous by its very nature}. \parencite[5, italics ours]{vonNeumann1932}
\end{quote}

It is noteworthy that in the massive, more recent literature on the foundations of quantum mechanics, the empirical input of the theory, as von Neumann, Bohr, Heisenberg, Born, Jordan, and Dirac, saw it, seems to have been entirely forgotten; we shall return to this point in \S3.

\item \label{correspondenceprincipleIni} Assumption II. \textit{The Correspondence Principle.}  Max Jammer writes \parencite[118]{jammer1966conceptual} that ``there was rarely in the history of physics a comprehensive theory that owed so much to one principle as quantum mechanics owed to Bohr's correspondence principle.'' This principle, which is genetically connected to assumption I,\footnote{See \cite{sep-bohr-correspondence} for a detailed discussion.} was a statement about how to determine, from the point of view of the framework of Bohr's model of the atom, quantum frequencies and intensities---i.e., probabilities of transitions between stationary states---from classical electrodynamics. Thus Darrigol writes:
\begin{quote}
Bohr assumed that, even for moderately excited states, the probability of a given quantum jump was approximately given by the intensity of the ‘corresponding’ harmonic component of the motion in the initial stationary state. This is what Bohr called ‘the correspondence principle’. \parencite[550]{Darrigol1997}
\end{quote}

For the convenience of the reader, let us now remark:
\begin{enumerate}
    \item The ``harmonic component of motion'' refers to the terms in a Fourier series expansion of the kinematic variable
    \begin{equation}\label{fourierexpansion}
    x(t)=\sum_{\alpha=-\infty}^\infty x_\alpha(n)\e^{\I \alpha \omega_n t}
\end{equation}
which is the solution of the mechanical equations that fix the stationary states of the system,\\
\noindent and, furthermore,
    \item in \eqref{fourierexpansion}, $\{\e^{\I \alpha \omega_n t}\}_{\alpha \in \mathbb{Z}}$ constitutes an orthonormal basis of the periodic functions with period $\nicefrac{2\pi}{\omega_n}$. The coefficients of the expansion, then---i.e., the amplitudes---, can thus be readily obtained in terms of the inner product
\begin{equation}\label{fouriercoeficient}
    x_\alpha(n)=\langle \e^{\I \alpha \omega_n t},x(t)\rangle:=\frac{\omega_n}{2\pi}\int_0^{\frac{2\pi}{\omega_n}} x(t)\e^{-\I  \alpha \omega_n t}dt.
\end{equation}
The quantities $|x_\alpha(n)|^2$ are the harmonic intensities. \label{correspondenceprincipleEnd}
\end{enumerate}
We now can, with \eqref{fourierexpansion} and \eqref{fouriercoeficient}, understand better how the principle of correspondence works: The key idea is that the harmonic components \eqref{fouriercoeficient}, calculated with the help of classical \textit{mechanics}, \textit{correspond to}---hence the name---the harmonics of the \textit{electric dipole moment} which, according to Lorentz's theory of the electron, determine the properties of the radiation that is emitted by an electron in motion (see \parencite[Chapter 2]{Bohr1923}).
\end{itemize} 

This fact cannot be overemphasized: The chief incentive to develop, in the beginning of the 20th century, a new mathematics to describe the new mechanics \textit{was the conviction that what distinguishes classical from quantum motion is that the latter is discontinuous}. Thus Poincaré wrote in 1912:

\begin{quote}
 It is hardly necessary to remark how much this concept [of quanta] differs from what we imagine to be true up to this point; physical phenomena would cease to obey the laws expressed as differential equations and this would undoubtedly be the greatest and the most profound revolution that nature has undergone since Newton. \parencite[5]{Poincare1912}   
\end{quote}

 With these preliminaries out of the way, we can now remark: \parencite{Heisenberg1925} is a step-by-step attempt to develop the needed discontinuous mathematics. This is why this paper came to be considered, by general acclaim, the birth of quantum \textit{mechanics}. 
 
Now it was apparent, from assumptions I and II, that one had to find the quantum-theoretical expressions for \eqref{fourierexpansion}, and therewith, for \eqref{fouriercoeficient} and $|x_\alpha(n)|^2$, since the motion of the atomic system that gives rise to the emission of radiation was now understood to be discontinuous (jumps between stationary states). How? Well, equation \eqref{fourierexpansion} is, of course, \textit{classical}, and, classically speaking, the intensity of radiation $|x_\alpha(n)|^2$ is proportional to $\nicefrac{dE}{dt}$. This suggests, if we think about it carefully, that, to find the quantum-theoretical expression for the intensity---or equivalently, \textit{the probability of a quantum jump}, for now we are interpreting the intensities of the emitted radiation as the \textit{frequencies with which the discontinuous transitions occur}---, the first order of business is to note that the energy of the radiation emitted in a quantum jump depends not on the time $t$, \textit{but on the frequency $\nu$}. For in view of this fact, given that $\nu = \nicefrac{\partial E}{\partial S_0}$, and given that the quantum expression for the variation of the energy is evidently \eqref{Bohrenergycondition}, all that was required, to write down the quantum-theoretical frequency $\nu(n', n)$---and therewith, a \textit{mechanics} based fundamentally on \textit{the observed emitted frequencies}---seems to be to write down \textit{the difference-expression for $S_{0}$}. We now emphatically remark: \textit{this}---$\Delta S_0$---is precisely \textit{what the canonical commutation-relations come down to}, in Matrix Mechanics, as the following remarks will demonstrate.

Having thus defined, with sufficient clarity, the problem Heisenberg dealt with in \parencite{Heisenberg1925}, let us now think through how he found the means to solve it. Since \eqref{Bohr-Sommerfeld} is an integral equation, and since one knows how to compute the derivative of an integral, to turn it into a \textit{difference}-equation one could take advantage of Born's 1924 calculus of differences. For Born, in 1924, in the attempt to develop the discontinuous mathematics to describe quantum jumps, considered (see \cite{Born1924}, \S2) a transition from a state labeled, say, by numbers $n_1, \ldots, n_f$ to a state labeled by $n_1+\tau_1, \ldots, n_f+\tau_f $, first, from the classical point of view. The transition takes place, of course, in this case, in a (continuous) \emph{linear way}, as Born refers to it, going through \text{all} the intermediary states $n_k+\mu \tau_k$, where $0\leq \mu \leq 1$. Next, Born considered the classical relation in the angle action variable $\nu_k=\nicefrac{\partial E}{\partial S_0{_k}}$ and the formula $(S_0)_k=h(n^\prime_k+\mu \tau_k)$, and obtained
\begin{equation}
    \tau \nu=\sum_{k} \tau_k \nu_k=\sum_k \tau_k \frac{\partial E}{\partial (S_0)_k}=\sum_k \frac{1}{h}\frac{d (S_0)_k}{d \mu} \frac{\partial E}{\partial (S_0)_k}=\frac{1}{h}\frac{dE}{d\mu}.
\end{equation}
Born's proposal was, then, this: Whenever a classical calculated quantity $\Phi$ yields
\begin{equation}\label{classicaldJII}
\sum_{k} \alpha_k \frac{\partial\Phi}{\partial (S_0)_k}=\frac{1}{h}\frac{d\Phi}{d\mu},
\end{equation}
the trick to transform it on a discontinuous (quantum), as opposed to a continuous (classical), quantity is to replace it by the difference coefficient as
\begin{equation}\label{quantumdJ}
\int_0^1\sum_{k} \alpha_k \frac{\partial\Phi}{\partial (S_0)_k}d\mu=\frac{1}{h}[\Phi(n+\alpha))-\Phi(n)].
\end{equation}

In particular, the expression of the emission of radiation in classical theory,
\begin{equation}\label{2indexfrequenceII}
    \nu(n, \alpha) = \alpha\nu(n)=\alpha \frac{\partial E}{\partial S_0}=\frac{\alpha}{h}\frac{dE}{dn}
\end{equation}
corresponds, in Born's scheme, to the quantum-theoretical expression 
 \begin{equation}\label{quantumfrequencyworkII}
    \nu(n, n-\alpha)=\frac{1}{h}(E(n) - E(n - \alpha)),
\end{equation}
which is indeed \eqref{Bohrenergycondition}. Thus Born, to construct quantum mechanics, helped himself to a kind of Wittgenstenian trick. He throws away the continuity ladder after he has climbed up it. 

A moment's reflection will reveal that the trouble with Born's procedure is that it was not general: one had to make the classical-to-quantum translation case-by-case. But the above considerations do exhaust ``Heisenberg's ideas on quantum conditions'' one needs to derive the canonical commutation relations; Born's calculus of differences is clearly built from the ground up on Assumptions I and II\footnote{For a discussion of Heisenberg's use of the correspondence principle in \parencite{Heisenberg1925} see, e.g., \parencite{MacKinnon1977} and \parencite[\S4.3]{sep-bohr-correspondence}.} and, from equation \eqref{fourierexpansion}, we note only that it naturally implies 
\begin{equation}\dot{x}(t)=\sum_{\alpha=-\infty}^\infty \I\alpha\omega_n x_\alpha(n)\e^{\I \alpha \omega_n t},\end{equation} so that we can define $p_\alpha(n)=\I\alpha m\omega_n x_\alpha(n)$ to have \begin{equation}m\dot{x}(t)=\sum_{\alpha=-\infty}^\infty p_\alpha(n)\e^{\I \alpha \omega_n t}.\end{equation} One last thing: Heisenberg also required that $x$ be real, so that $x_\alpha(n)=\overline{x_{-\alpha}(n)}$. We can now reconstruct Born's reasoning, the way he ``condensed Heisenberg's ideas on quantum conditions,'' by applying all of this in the classical expression of the reduced action $S_0$ to find therewith its quantum-theoretical counterpart.

\begin{align}
S_0&=\oint p\dot{q} d\gamma\\
&=m\int_0^T \dot{x}(t)^2 dt\\
&=-m\sum_{\alpha,k=-\infty}^\infty \alpha\omega_n x_\alpha(n) k\omega_n x_k(n) \int_0^{\frac{2\pi}{\omega_n}}  \e^{\I (\alpha+k) \omega_n t} dt\\
&= 2\pi m\sum_{k=-\infty}^\infty k^2\omega_n x_k(n) x_{-k}(n)\\
&= 2\pi \I \sum_{k=-\infty}^\infty k \left(x_k(n)\e^{\I k\omega_n t}\right) \left(i(-k)m\omega_n x_{-k}(n)\e^{\I (-k)\omega_n t} \right)\\
&= 2\pi \I \sum_{k=-\infty}^\infty k \left(x_k(n)\e^{\I k\omega_n t}\right) \left(p_{-k}(n)\e^{\I (-k)\omega_n t} \right),
\end{align}
where the integration, as prescribed by the old quantum theory, was performed over the closed curve corresponding to one period of the harmonic components of motion (see equation \eqref{fourierexpansion}), namely, $T=\nicefrac{2\pi}{\omega}$---and,  fortuitously, \begin{equation}\int_0^{\nicefrac{2\pi}{\omega_n}}  \e^{\I (\alpha+k) \omega_n t} dt=\nicefrac{2\pi}{\omega_n} \delta_{\alpha,-k}.\end{equation}
Taking the derivative in the quantity $S_0$ yields
\begin{align}
1= 2\pi \I \sum_{k=-\infty}^\infty k \frac{\partial}{\partial S_0}\left(\left(x_k(n)\e^{\I k\omega_n t}\right) \left(p_{-k}(n)\e^{\I (-k)\omega_n t} \right)\right).
\end{align}

So far, everything is still classical. To make it quantum-theoretical, we use Born's relation \eqref{quantumdJ}, to get

\begin{align}\label{classicalCCRII}
1= \frac{2\pi \I}{h} \sum_{k=-\infty}^\infty \left(x_k(n+k)\e^{\I k\omega_{n+k} t}\right) \left(p_{-k}(n+k)\e^{\I (-k)\omega_{n+k} t} \right)-\left(x_k(n)\e^{\I k\omega_n t}\right) \left(p_{-k}(n)\e^{\I (-k)\omega_n t} \right).
\end{align}
But the problem is clearly not yet solved, for the \textit{frequencies} in \eqref{classicalCCRII} are not in agreement with \eqref{Bohrenergycondition}. We now categorically make the statement: \emph{it is to correct for this that Heisenberg rewrote the equations in two indices}. The harmonic frequencies can easily be expressed in two indices, but coherence, \textit{and the correspondence principle}, demands that the \text{amplitudes} $x_k(n)$ be replaced by $x(n,k)$. The problem was how to express sums of \textit{products} of such coefficients; the solution, Heisenberg noted, was not far to seek, for conservation of energy clearly implies, for two consecutive energy transitions in Bohr's atomic model, that
\begin{equation}
    \nu(n,\alpha)+\nu(n, \beta)=\nu(n,\alpha+\beta) \quad \longrightarrow \quad \nu(n,n-\alpha)+\nu(n-\alpha,n-\alpha -\beta)=\nu(n,n-\alpha-\beta).
\end{equation} 
If we then use \eqref{fourierexpansion} to formally analyse the expression of $x(t)^2$, we get
\begin{align}
    x(t)^2&:=\left(\sum_{\alpha=-\infty}^\infty x_\alpha(n)\e^{\I \alpha \omega_n t}\right)\left(\sum_{\gamma=-\infty}^\infty x_\gamma(n)\e^{\I \gamma \omega_n t}\right)\\
    &=\sum_{\beta=-\infty}^\infty\sum_{\alpha=-\infty}^\infty x_\alpha(n) x_\gamma(n) \e^{\I (\alpha+\gamma)\omega_n t}\\
    &=\sum_{\gamma=-\infty}^\infty \left(\sum_{\alpha=-\infty}^\infty x_\alpha(n) x_{\beta-\alpha}(n)\right) \e^{\I \beta\omega_n t},
\end{align} 
where we have renamed the index $\beta:=\alpha+\gamma$ to isolate the harmonic components. For a product $c$ of two quantities $a$ and $b$ we can then infer, like Heisenberg did,\footnote{Note that the validity of the general formula \eqref{eq:productcoeficients} follows from considering the product $x(t)y(t)$ of two arbitrary quantities expanded in Fourier coefficients.} that the rule should be indeed
\begin{equation}\label{eq:productcoeficients}
  c_\beta(n)=\sum_{\alpha} a_{\alpha}(n)b_{\alpha-\beta}(n) \quad \longrightarrow \quad   c(n,n-\beta)=\sum_{\alpha} a(n,n-\alpha)b(n-\alpha,n-\beta).
\end{equation}
Applying this to \eqref{classicalCCRII} readily yields 
\begin{equation}\label{preCCRII}
1= \frac{2\pi \I}{h} \sum_{k=-\infty}^\infty \left(x(n+k,n)\e^{\I \omega(n+k,n) t}\right) \left(p(n,n+k)\e^{\I \omega(n,n+k) t} \right)-\left(x(n,n-k)\e^{\I \omega(n,n-k) t}\right) \left(p(n-k,n)\e^{\I \omega(n-k,n) t} \right).
\end{equation}

\noindent Recalling now that the product of two (infinite, by abuse) matrices $A=(a_{ij})_{i,j=1}^\infty$ and $B=(b_{ij})_{i,j=1}^\infty$ is given by $(AB)_{ij}=\sum_{k=1}^\infty a_{ik} b_{kj}$, to Born is due the credit of recognizing that the objects above behave, as far as multiplication is concerned at least, like matrices (see \parencite{Born1955} and \parencite{mehra1982historical})---and therewith the realization that one needed not invent a new mathematics to deal with quantum phenomena after all. One could use the fully developed matrix calculus. Carrying out the identification, change of summation-index leads directly to the conclusion that the above expression reduces to
\begin{equation}
    -\I  \hbar=(px-xp)_{ii}, \quad \forall i\in \mathbb{N}.
\end{equation}

This proves our contention that the canonical commutation relations amount to a kind of $\Delta S_0$. The claim that it necessarily encodes discontinuity, as emphasized by \parencite{heisenberg1927}, is, however, more complex than meets the eye, as the results of the next section shall unequivocally demonstrate. 

 \subsection{The Dynamical Origin of the Quantization Rules, Part II: Wave Mechanics}

It has been taken for granted, in the history and philosophy of physics literature, that the canonical commutation relations, in Wave Mechanics, have played no role whatever prior to Schrödinger's equivalence proof.\footnote{\cite{jammer1966conceptual}; \cite{mehra1987erwin}; \cite{Mullermyth1}; \cite{MULLERMYTH2}; \cite{beller1999quantum}; \cite{bitbol2012schrödinger}; \cite{PEROVIC2008}.} The standard view is that the commutation relations were \textit{posited} by Schrödinger \textit{to obtain equivalence} with Matrix Mechanics; ``Schrödinger only invented these [canonical] operators,'' Muller writes, for example, in his sophisticated, and widely read, analysis of the mathematical isomorphism between Matrix and Wave Mechanics, ``when pursuing his equivalence proof'' \parencite[46]{Mullermyth1}. A more detailed mathematical analysis of Schrödinger's first paper on Wave Mechanics will however reveal that this is not the case; in fact, the canonical commutation relations are a direct consequence of Schrödinger's \textit{first} derivation of the wave equation (but---most emphatically---not the second), and so have been part and parcel of Wave Mechanics since its inception. This result, to be demonstrated in this section, will lead us to ask questions about the relationship between Matrix and Wave Mechanics that Muller did not consider in his influential investigation---and therewith, instructive facts involved in the equivalence that have been hitherto hidden from sight will naturally emerge.\footnote{ \label{foot:11}The fact that the canonical commutation relations is entailed by the first, but not by the second, derivation of the wave equation will entail, as we shall later see, that only the first---but again, not the second---derivation of the wave equation is connected to the Matrix Mechanics formalism. This fact, brought about by our mathematical analysis, has an important consequence for discussions about the interpretation of quantum mechanics, for it immediately leads to the solution of an otherwise perplexing historical problem that foundational discussions cannot afford to overlook: if it is admitted, as it now generally is, that theoretical formalisms carry their ontologies within themselves, so that the different ``interpretations'' of quantum mechanics are in point of fact different theories, then how it can \textit{be} that the formalisms of Matrix and Wave Mechanics were taken to be equivalent, and yet, their authors disagreed about how to interpret it? The reason is this: It follows from the fact that Schrödinger's first and second derivations depend on entirely different premises that they must be logically and conceptually and physically \textit{distinguished} from one another.  The physics, and hence the ontology, encoded in the wave equation in \parencite{schrproper1} and \parencite{schrproper2} are indeed not the same.  And so it will now further follow, from the fact, to be proven later, that Schrödinger's \textit{first} derivation connects Matrix and Wave Mechanics (he used this derivation in his equivalence proof), together with the fact that it is \textit{on the second} derivation of the wave equation that Schrödinger's ``interpretation'' of quantum mechanics depends (see, e.g., \parencite{MackKinnon1980, Wessels1980} ), that Schrödinger's, and the statistical, interpretation \textit{are indeed attached to different formalisms}. In this way, the conceptual ambiguity between theory and interpretation that the equivalence between Matrix and Wave Mechanics otherwise inexorably entails is completely resolved. }

The oversight mentioned in the previous paragraph is related to the fact there is a certain prejudice, in the scholarship on Wave Mechanics, to think of Schrödinger's first derivation of the wave-equation \parencite{schrproper1} as having only a formal significance, as it seems, on a first and second glance, to depend on seemingly physically-unmotivated steps (\cite{WESSELS1979}; \cite{Wessels1980}; \cite{MackKinnon1980}; \cite{Kragh1982-KRAESA-6} \cite{Mullermyth1}). We will try to remove this prejudice by proceeding here as in the previous section, and doing for Schrödinger what Schrödinger didn't do for himself: we shall be absolutely explicit about what premises we are using, and why. 

Consider then, by way of preamble, the opening remarks of Schrödinger's first paper on Wave Mechanics. He is warning his readers---though from a modern perspective, this may not be obvious---, that he is playing a different kind of game than that of the matrix theoreticians. 

\begin{quote}
In this communication I would like first to show, in the simplest case of the (non relativistic and \textit{unperturbed}) hydrogen atom, that the usual prescription for quantisation can be substituted by another requirement in which no word about “integer numbers” occurs anymore. Rather, the integerness emerges in the same natural way as, for example, the integerness of the number of knots of a vibrating string. The new interpretation is generalisable and touches, I believe, very deeply the true essence of the quantisation prescription. \parencite[1, italics ours]{schrproper1}
\end{quote}

This quote, and the discussion in \S 2.1, tell us an exceedingly important fact, namely that while Matrix Mechanics was an attempt to improve, and generalize, the old quantum theory associated with the names of Planck, Einstein, and Bohr, \textit{Wave Mechanics was an attempt to overcome it}. Schrödinger wanted to demonstrate, by direct counter-example, that the discontinuity hypothesis should by no means be taken as ``empirically given.'' 

To proceed, then, to the argument: the reader will remember that in the old quantum theory the quantum conditions, i.e., the Bohr-Sommerfeld rules, are posited over the (reduced) Hamilton-Jacobi equation
\begin{equation}\label{eq:HamiltonJacobiII}
    H\left(q,\frac{\partial S}{\partial q}\right)=E.
\end{equation}
Schrödinger too will make use of the Hamilton-Jacobi formulation of mechanics. The difference is that he will not postulated over them, but rather---through the introduction of a new fundamental hypothesis he did not himself, but we will, properly discuss---, \textit{elicit from them}, the allowed quantum-mechanical motions.

The procedure is as follows. Start by substituting the action S with a new unknown $\psi$\footnote{Following Schrödinger, we will consider $\psi$ to be everywhere real function, but we remark that one can complexify the space and sesqui-linearly extend the inner product to get the corresponding quadratic form \eqref{eq:functionalJII}.} via the transformation
\begin{equation}\label{eq:PCFTranfII}
S=K\log \psi, \quad K\in \mathbb{R}_+.
\end{equation}
Inserting the latter into \eqref{eq:HamiltonJacobiII} yields
\begin{equation}\label{eq:HJpsiII}
H\left(q,\frac{K}{\psi}\frac{\partial \psi}{\partial q}\right)-E = \frac{K^2}{2m}\psi^{-2}\left(\frac{\partial \psi}{\partial q}\right)^2+V(q)-E=0. \quad \footnote{Here we are adopting the notation $\nicefrac{\partial \psi}{\partial q}=\operatorname{grad}(\psi)$, $\left(\nicefrac{\partial \psi}{\partial q}\right)^2=\left(\nicefrac{\partial \psi}{\partial q}\right)\cdot \left(\nicefrac{\partial \psi}{\partial q}\right)$, and $\nicefrac{\partial^2 \psi}{\partial q^2}=\nabla^2 \psi$.}\end{equation}

Now evidently, solving \eqref{eq:HJpsiII} \textit{as it stands} is merely to solve the classical equations of motion in a new variable. It doesn't change anything. And it is important to now remember that the Hamilton-Jacobi equations, in classical physics, describe \textit{only} the motion of \textit{localized} objects, i.e., point-particles and/or wave-packets. Now Schrödinger, of course, did not merely solve the classical equations of motion in a new variable. Rather he sought to define a \emph{variational problem} in $\psi$.

Only one must note what \textit{that} means and entails, in conceptual, and specifically in \textit{dynamical}, terms.

But before we can get into interpretative questions let us carry through the calculation formally. We follow Schrödinger, and multiply the left-hand side of equation \eqref{eq:HJpsiII} by $\psi^2$, and then integrate it in space, thereby defining the quadratic form 
\begin{equation}\label{eq:functionalJII}
\CJ[\psi]=\int_{\mathbb{R}^3} \frac{K^2}{2m}\left(\frac{\partial \psi}{\partial q}\right)^2+V(q)\psi^2-E\psi^2 dq=\lim_{R\to \mathbb{R}^3}\int_{\partial R} \frac{K^2}{2m}\psi\frac{\partial \psi}{\partial q}  d\vec{S}+\int_{\mathbb{R}^3} -\frac{K^2}{2m}\psi\frac{\partial^2 \psi}{\partial q^2}+V\psi^2-E\psi^2 dq. \quad \footnote{Later in \parencite{schrproper1} Schrödinger added a regularity condition that will require the first integral of the right-hand side to vanish.}
\end{equation}
The variational problem is then defined by requiring not that $\CJ[\psi]$ vanishes, as suggested by equation \eqref{eq:HJpsiII}---and from which, let us remember, the classical equations of motion for localized objects can be derived. Instead we must impose that the functional $\CJ$ be \emph{stationary} at $\psi$:
\begin{equation}\label{SchrMinII}
\frac{1}{2}\delta \CJ[\psi]=\lim_{R\to \mathbb{R}^3}\int_{\partial R} \frac{K^2}{2m}\frac{\partial \psi(q)}{\partial q} \delta\psi(q) d\vec{S}+\int_{\mathbb{R}^3} \left(-\frac{K^2}{2m}\frac{\partial^2 \psi(q)}{\partial q^2}+V(q)\psi(q)-E\psi(q)\right)\delta\psi(q) dq=0.
\end{equation}

 A key passage of \parencite{schrproper1} is Schrödinger's emphasis that the quantum postulate of the old quantum theory, \emph{i.e}, the Bohr-Sommerfeld quantization rules, should be \textit{replaced} by this variational problem. The arbitrariness of $\delta \Psi$ does indeed imply that the terms in the integrals must vanish; the first term in \eqref{SchrMinII} is a boundary condition, upon which Schrödinger imposes regularity conditions that imply that it too must vanish. Hence the variational problem is reduced to the second term on the right-hand side,
\begin{equation}\label{eq:SchrII}
-\frac{K^2}{2m}\frac{\partial^2 \psi(q)}{\partial q^2}+V(q)\psi(q)-E\psi(q)=0,
\end{equation}
which, as the reader shall immediately recognize, is the time-independent Schrödinger equation.

Now to the question of the meaning of the transition from \eqref{eq:HJpsiII} to \eqref{eq:functionalJII}. We want to bring to the open two facts that are hidden in it. 

(The move to \eqref{SchrMinII}, which outstanding scholars have hitherto deemed ``cryptic'' and ``\textit{ad hoc}'' (\cite{WESSELS1979}; \cite{Kragh1982-KRAESA-6}; \cite{JOAS2009}), necessitates more careful mathematical analysis---we shall be able to demystify it in \S 2.4.)

\begin{enumerate}
    \item \textbf{The nature of the classical limit.} The point cannot be overemphasized: the Hamilton-Jacobi equation \eqref{eq:HamiltonJacobiII} describes the motion of localized objects, and replacing $S$ by $\psi$ in it, via transformation \eqref{eq:PCFTranfII}, does not change this fact. The physically important question is therefore what is the effect of transforming equation \eqref{eq:HJpsiII} into \eqref{eq:functionalJII}; the answer is that this transition forces the integrand to be seen as a Hamiltonian \emph{density}, from which fact it follows that $\psi$ must now necessarily and ineluctably acquire the significance of a \emph{continuous quantity distributed in space}.
    
    To accept Schrödinger's \parencite{schrproper1} derivation of the wave equation thus forces us to acknowledge that the passage from quantum to classical motion is the passage of showing just how this continuously-distributed quantity $\psi$ behaves, in the appropriate classical limit, like localized wave-packets. The fact that this information is ``hidden'' in the mathematics deserves comment: Schrödinger was only explicit about the Hamiltonian analogy between Mechanics and Optics---which implies the fact just stated, namely that classical mechanics is a limiting case of a more general wave mechanics---in \parencite{schrproper2}, his second communication on Wave Mechanics; historians have interpreted the absence of an explicit reference to the Hamiltonian analogy in \parencite{schrproper1} as evidence of the fact that Schrödinger didn't even know, or, didn't fully understand, the optical-mechanical analogy when he started the quantization series (see \parencite[\S4]{JOAS2009}). Our analysis, however, indicates otherwise. Schrödinger ought to have known what he was doing, and the mathematics of \parencite{schrproper1} is unequivocal: It is an exact mathematical transcription of the optical-mechanical analogy. Schrödinger's \parencite{schrproper1} quantization procedure not merely allows, but indeed \textit{demands}, as a matter of simple self-consistency, that the classical limit be recovered in the way just described.\footnote{The matrix theoreticians used the fact that the classical limit is not recovered in this way as a strong argument against the general consistency of Schrödinger's position---and to great effect; failure to solve this and other related difficulties soon led to the downfall of Schrödinger's ``interpretation.'' See the details in \parencite[\S6]{MackKinnon1980} and \parencite[chapter 2]{beller1999quantum}.} 

    \item \textbf{The quantum conditions}. The passage from \eqref{eq:HJpsiII} to \eqref{eq:functionalJII} also requires that the quadratic form on the second part of the integrand on the right-hand side of \eqref{eq:functionalJII} must be the result of the action of an operator\footnote{In this case the operator that defines the quadratic form in terms of the inner product is obvious, but this is in fact the general case: By standard techniques, one can construct a sesqui-linear form from which the quadratic form $Q$ is obtained. Then, thanks to the unbounded version of the Riesz Representation theorem for the densely defined, symmetric, and closed sesqui-linear form, one obtains a unique densely defined closed unbounded operator $A:D(A)\subset \mathcal{H}\to \mathcal{H}$ such that $Q(\psi)=\ip{\psi}{A\psi}$.} on $\psi$:
$$(\hat{H}(q,p)-E)(\psi)(q)=-\frac{K^2}{2m}\frac{\partial^2 \psi(q)}{\partial q^2}+(V(q)-E)\psi(q),$$
which fact, added to the requirement that the procedure must hold for an arbitrary potential $V(q)$, implies in identifying $p$ and $q$, \textit{via} correspondence with \eqref{eq:HJpsiII}, with operators defined by $(\hat{p} \psi)(q)=\pm i K \nicefrac{\partial \psi}{\partial q}(q)$ and $(\hat{q} \psi)(q)=q\psi(q)$. The identification Schrödinger spelled out only later, in \parencite[\S2]{schequivalence}---namely, that the ``$p_l$ in the [classical] function is to be replaced by the operator $\nicefrac{\partial}{\partial q_l}$'' \parencite[47]{schequivalence}---, is therefore obtained automatically, in his first communication on Wave Mechanics, by the transition from \eqref{eq:HJpsiII} to \eqref{eq:functionalJII}. \footnote{It must be emphasized here that flagrant contradictions between \eqref{eq:PCFTranfII} and the classical (abbreviated) action $S_0=\int pdq$ inevitably follow if the mathematical significance of the transition from \eqref{eq:HJpsiII} to \eqref{eq:functionalJII} is overlooked; see  \cite{SkalaKapsa2007}. This is manifested in the fact that is not a coincidence, but instead a mathematical requirement, that the imaginary unit is not incorporated in the constant $K$, but appears only in the definition of $\hat{p}$. The discussion about the correct interpretation of the abbreviated action $S_0$, which takes into account the momentum as an operator, will be done in \S 2.4, equation \eqref{eq:nick}, below.} In \parencite[\S2]{schrproper1}, Schrödinger remarked that experiments require that we set $K=\nicefrac{h}{2\pi}$. It must therefore be acknowledged---and this is our main point---that the transition from \eqref{eq:HJpsiII} to \eqref{eq:functionalJII} also inexorably and ineluctably entails that $[\hat{p},\hat{q}]=-\I  \hbar$.
\end{enumerate}
Our assertion in the opening paragraph of this section---namely, that Schrödinger's first derivation of the wave equation \textit{implies} the canonical commutation relations---has thus been proved.

\subsection{The Dynamical Origin of the Quantization Rules, Part III: The Problem of the ``Inner Connection'' Between Matrix and Wave Mechanics }

Equipped with the results of the two previous sections, we now wish to consider the opening lines of \parencite{schequivalence}. This is Schrödinger's so-called ``equivalence paper.'' 

\begin{quote}
    Considering the extraordinary differences between the starting-points and the concepts of Heisenberg’s quantum mechanics and of the theory which has been designated “undulatory” or “physical” mechanics, and has lately been described here, it is very strange that these new two theories \textit{agree with one another with regard to the known facts}, where they differ from the old quantum theory. I refer, in particular, to the peculiar “half-integralness” which arises in connection with the oscillator and the rotator. That is really very remarkable because starting-points, presentations, methods and in fact the whole mathematical apparatus, seem fundamentally different. [...] Above all, however, the departure from classical mechanics in the two theories seems to occur in diametrically opposed directions. In Heisenberg’s work the classical continuous variables are replaced by systems of \textit{discrete} numerical quantities (matrices), which depend on a pair of integral indices, and are defined by \textit{algebraic} equations. The authors themselves describe the theory as a “true theory of a discontinuum”. On the other hand, wave mechanics shows just the reverse tendency; it is a step from classical point-mechanics towards a \textit{continuum}-theory. In place of a process described in terms of a finite number of dependent variables occurring in a finite number of differential equations, we have a continuous \textit{field-like} process in configuration space, which is governed by a single partial differential equation, derived from a Principle of [Least] Action.
\end{quote}

The fact to be explained, as Schrödinger sees it, is therefore this: ``the two theories agree [numerically] with one another with respect to the known facts.'' There are four points to be made in this connection. 

The first is that since Muller's persuasive \parencite{Mullermyth1, MULLERMYTH2} it has been generally taken for granted by the literature that Schrödinger didn't prove that the canonical matrix- and wave-operator algebras are isomorphic. We will proceed to question the mathematical correctness of Muller's disproof of Schrödinger's isomorphism proof in a separate paper; here, we need not, and so will not, get into it. For with or without a rigorous proof of algebra isomorphism---and this is our second point---, the further fact remains:
\begin{quote}
[The] relation of matrices to functions is \textit{general}; it takes no account of the \textit{special} system considered, but is the same for all mechanical systems. (In other words: the particular Hamiltonian function does not enter the connecting law.) \parencite[46]{schequivalence}
\end{quote} 

In other words, \textit{kinematical equivalence does not entail dynamical equivalence.} This fact does not emerge with full clarity in Muller's investigation, as he focuses in the isomorphism: that dynamical equivalence between Matrix and Wave Mechanics must be proved \textit{independently} from the isomorphism between matrices and wave-operators (this fact shall be important for us in \S3.1). This is why Schrödinger sets out to prove, in \parencite[\S4]{schequivalence}, that to solve his minimization problem \eqref{eq:SchrII} is equivalent to solving the Matrix Mechanics problem of diagonalizing the matrix $H$. 

However, let us reflect. It will quickly appear that this argument alone cannot be taken as the ground of dynamical equivalence. For the diagonalization of the matrix $H$ provides no indication of how the quantum system evolves in time; the dynamical posits of each theory may well be different.  Indeed, if by the time Schrödinger wrote the equivalence paper (March 1926), Matrix and Wave Mechanics did not postulate the exact same dynamical laws, then it trivially follows that the theories were not equivalent. And since Schrödinger's treatment of the transitions (intensity lines) appeared only \textit{after} the equivalence paper, in \parencite{schrproper3}, wherein he explicitly says that he \textit{imported the rule}, \textit{via} the mapping between matrix-coefficients and wave-functions described in the equivalence proof (see \parencite[\S4]{schrproper3}), \textit{from} Matrix Mechanics (this fact shall also be important for us in \S3.1.), we find, ineed, \textit{in this fact alone}, conclusive evidence that establishes that by the time Schrödinger wrote the equivalence proof, they indeed were not.

The third point we want to make is that the way Schrödinger related his theory to Heisenberg, Born and Jordan's falls short of explaining their connection. For Schrödinger explicitly remarked, let us repeat it expressly once more, that matrix- and wave-operators isomorphism does not entail theoretical equivalence; furthermore, by the argument we have just given, the theories were indeed, as of March 1926, neither mathematically nor empirically equivalent. But the puzzle of how it can then \textit{be} that they ``agree [numerically] with one another with respect to the known facts” apparently remains. And it is against this fact that we now ask the reader to note, and this is our fourth and final point, that, by the lights of our discussion, the numerical agreement of Matrix Mechanics and Wave Mechanics for simple toy systems that depend only on $p$ and $q$ such as those Schrödinger mentions is \textit{not} strange. It is an unavoidable mathematical consequence of the fact that \textit{both theories carry the canonical commutation relations in themselves.} 

For to say that Matrix and Wave Mechanics agree in that $[\hat{p}, \hat{q}] = -\I   \hbar$ \textit{because} there is a mathematical isomorphism between them is, please note, \textit{to invert the arrow of explanation}. For it is $[\hat{p}, \hat{q}] = -\I   \hbar$ that characterizes the algebra. It is therefore the fact that $[\hat{p}, \hat{q}] = -\I   \hbar$ holds for both theories that will allow an isomorphism between them to be established---not the other way around.  

And so it will now follow that it is \textit{this}, and \textit{not} the numerical agreement itself, as Schrödinger thought, that is the fact to be explained. The fact that theories that imply and are implied by apparently mutually contradictory hypotheses both \textit{entail}  $[\hat{p}, \hat{q}] = -\I   \hbar$ is a truly astonishing fact that must itself have a deeper significance. It suggests that there is \textit{something} here, a deeper, \textit{physical} fact, a common underlying structure, that supports or necessitates the link between Matrix and Wave Mechanics---the link upon which von Neumann erected the mathematical structure of orthodox quantum mechanics (we shall discuss this point in detail in \S3). The purpose of the next section is to bring out this deeper and effectively forgotten matter-of-fact finally to the surface.

\subsection{The Dynamical Origin of the Quantization Rules, Part IV: The Dual Action}
\label{sec:duality}

Now if it is, as we have remarked above, a common error to regard Schrödinger's first derivation of the wave equation as ``cryptic'' and ``\textit{ad hoc},'' this, in part, is because Schrödinger encouraged the mistake himself: In the opening lines of his second paper of Wave Mechanics, \parencite{schrproper2}, wherein a different, ``more intuitive,'' derivation of the wave equation is given, Schrödinger, in referring to the first paper, remarked that the transition from \emph{the equating to zero} of a certain expression to the postulation that the \emph{space integral} of the said expression shall be \emph{stationary}---i.e., the passage from \eqref{eq:HJpsiII} to \eqref{SchrMinII}, the very construction of his variation problem---is in itself incompreensible \parencite[13]{schrproper2}. But let us not isolate the word ``incomprehensible'' from the qualifier ``in itself.'' A principled justification for Schrödinger's variational problem must exist, even though he did not state it. And it will be our task to find it.

For we cannot afford to bypass this historical/mathematical/conceptual difficulty in our study. The relation $[\hat{p}, \hat{q}] = -\I   \hbar$ follows, as a matter of mathematical necessity, in Wave Mechanics only from the apparently ``cryptic,'' ``in itself incomprehensible,'' derivation of the wave equation given in \parencite{schrproper1}---it doesn't follow from the ``intuitive'' derivation given in \parencite{schrproper2}.\footnote{\label{foot:equivalencepaper}Note indeed that in the passage from the equivalence paper quoted above Schrödinger refers to the first derivation of the wave-equation, and not the second, as ``the'' derivation of the wave equation. This is in good agreement with the fact that the first derivation---and not the second, let us reiterate---implies the canonical commutation relations.} \textit{Nothing in the latter paper implies identifying classical functions with operators}. And since we can expect that the fact that necessitates the emergence of the commutation relations in Wave Mechanics is mathematically and conceptually \textit{tied}, by the upshot of the result of the previous section, to the fact that necessitates the emergence of the commutation relations in \textit{Matrix Mechanics}, it appears that we must only find the justification for Schrödinger's ``incomprehensible'' transition to find the key to the solution of the problem we have posed at the end of the previous section. And this, as the reader shall soon see, will turn out to be exactly the case.

We start by thinking through the assumptions involved in the derivation discussed in \S2.2 in a bit more detail. Schrödinger's first ingredient, let us remember, is the reduced Hamilton-Jacobi equation. A word about the Hamilton-Jacobi Theory: In the Hamiltonian formalism, the Hamilton-Jacobi equation is obtained by a canonical transformation that makes the transformed Hamiltonian vanish (the transformed action must still satisfy Hamilton's principle, of course). The \textit{reduced} Hamilton-Jacobi equation is a particular case of this, which appears when (a) the Hamiltonian does not depend on time, or (b) one proceeds by additive separation of variables. A few pages back we saw that the passage from \eqref{eq:HJpsiII} into \eqref{eq:functionalJII} involves replacing functions by functionals. The question comes naturally, then, as to whether the form of the Hamilton-Jacobi equation should not be adapted accordingly. So let us verify what follows from doing precisely this: imposing Hamilton's principle and separation of variables upon replacing functions by functionals in the new action function \eqref{eq:PCFTranfII}.

Let's start slow. The quantity action, by definition,
\begin{align}\label{eq:S=WHTII}
S=\int L dt=\int p\dot{q} dt-\int H dt=S_0-\int H dt,
\end{align}
where $S_0:=\int p\dot{q} dt$ is the reduced action, and $H$ is the Hamiltonian, encodes, let us remember, the change of a physical system over time. The equations of motion of a physical system are then derivable by the postulation that the action be stationary. If one is in the particular case, however, as Schrödinger indeed was, wherein one has to proceed by separation of variables (for Schrödinger was assuming conservation of energy), then the quantity $S$ reads instead
\begin{align}\label{eq:S-sepvariaveisII}
S=\int L dt=S_0-Et,
\end{align}
where $E$ is a constant. In the standard formulation of Hamiltonian mechanics, equations \eqref{eq:S=WHTII} and \eqref{eq:S-sepvariaveisII} lead to $H(q(t),p(t))=E$; here, we must adapt it. We must follow out the fact that the transition from \eqref{eq:HJpsiII} to \eqref{eq:functionalJII} forces us to reinterpret such equations in terms of functionals. We then have the replacement rule

\begin{align}\label{SepH-JfunctionalII}
S&=S_0-\int H dt \hspace{-3.5cm}&\leftrightarrow \quad \CS&=\CS_0-\int \tilde{\CJ} dt, \quad \textrm{and} \\
S&=S_0-Et \hspace{-3.5cm}&\leftrightarrow \quad \CS&=\CS_0-\int \tilde{\CE} dt,
\end{align}
where
\begin{align}
\label{eq:functionalJIII}\tilde{\CJ}[\psi]&=\int_{\mathbb{R}^3}-\frac{K^2}{2m}\psi(q)\frac{\partial^2 \psi(q)}{\partial q^2}+V(q)\psi(q)^2dq, \quad \textrm{and}\\
\tilde{\CE}[\psi]&=\int_{\mathbb{R}^3}E\psi(q)^2dq.
\end{align}

This makes it clear that it is an immediate consequence of equations \eqref{SepH-JfunctionalII}, and Hamilton's principle \textit{alone}, that
\begin{equation}\label{eq:functionalH-JII}
\delta \CS_0-\delta\left(\int \tilde{\CJ} dt\right)=\delta S=0=\delta \CS_0-\delta\left(\int \tilde{\CE} dt\right)\Rightarrow \delta\left(\int (\tilde{\CJ}-\tilde{\CE}) dt\right)=0\Rightarrow \delta \CJ=\delta(\tilde{\CJ}-\tilde{\CE})=0.\footnote{In the last implication we used that \eqref{eq:functionalJIII} does not depend on $t$. We stress that it is a consequence of separation of variables previously used and the fact that the Hamiltonian does not depend explicitly on $t$.}
\end{equation}

This is exactly Schrödinger's variational problem. Though Schrödinger called it ``incomprehensible,'' once the steps are explained, nothing can seem more justified. The stationary condition \eqref{SchrMinII}, or equivalently, the Schrödinger equation, \textit{is a logical consequence of applying Hamilton's principle to the action written in terms of the continuously spatially-distributed quantity $\psi$}.

The reader may wonder if this procedure does not amount to finding an action for a field. We categorically assert: it does not. For there is no direct generalization of the theory of canonical transformations for fields, and hence no counterpart of the Hamilton-Jacobi equation in field theory.\footnote{The Hamiltonian formulation for classical fields requires an infinite-dimensional manifold $M$ and the introduction of functional derivatives. This is why it is much harder to establish a symplectic manifold structure, and therefore, a theory of canonical transformations, in the cotangent bundle $T^*M$. For more details, we refer to \parencite{abraham2008}, in particular, it is worth checking remarks (3) and (4) on p. 383.} Recall, indeed, from \S 2.2: it is a peculiarity of \textit{this} procedure of Schrödinger's that it involves ``transforming'' the quantity $\psi$ from a discrete to a continuous object (optical-mechanical analogy). The elegance of it is undeniable; one must agree with Born: ``it would have been beautiful if you [Schrödinger] were right'' (Born to Schrödinger, 6 November 1926, quoted in \parencite[36]{beller1999quantum}). For it is precisely \textit{this} transformation, namely the transformation that does the work of the optical-mechanical analogy, that ``quantizes'' the system.

This completes our analysis of how the canonical commutation relations are implied by Schrödinger's derivation of the wave equation from a principle of Least Action. If we were right in expecting that the facts that lead to the emergence of the canonical commutation relations in Matrix Mechanics, and the facts that lead to the emergence of the canonical commutation relations in Wave Mechanics, are theoretically linked, then we should now be able to identify the link exactly. That is exactly the case---consider equations \eqref{SepH-JfunctionalII} and \eqref{eq:functionalH-JII} again.  

Schrödinger focused on the \textit{second} term of the action; he promoted $H$ to $\tilde{\CJ}$ and, as we just saw, the condition $\delta \tilde{\CJ}[\psi] = 0$ is equivalent to Schrödinger's equation. Consistency, however, suggests---perhaps demands---that the first term of the action also be considered. And all the detailed historical and technical analyses we have carried out up to now in this paper have finally served their purpose by enabling the reader to recall at once that the matrix theorists derived the canonical commutation relations starting from $\int p\dot{q} dt$, which is just $S_0$. And that is precisely the first term of the action. 

We thus now categorically conjecture the following: A \textit{duality} between Matrix and Wave Mechanics, \textit{grounded on the action}, must hold. To prove it we shall extend Schrödinger's treatment and apply it to the first term of the action, defining thereby the functional $S_0\leftrightarrow \CS_0$. The canonical commutation relations will be shown to follow provided $\CS_0$ satisfies a remarkably important physical condition.\\

We begin by noting that if one decides to follow Schrödinger's \parencite{schrproper1} procedure naively,\footnote{We are still following Schrödinger's assumption that the wave-function must be real. The case of complex wave functions will become clear a few steps later.} then, as we saw in \S 2.2, the replacement rule $p\rightarrow\hat{p}\psi:= -\I  K\nicefrac{\partial \psi}{\partial q}$ and $q\rightarrow \hat{q}\psi:=q\psi$ must hold. But there is an ambiguity here regarding $\dot{q}=\nicefrac{1}{m} p$; we shall assume that the correct replacement rule is $\dot{q}\rightarrow \nicefrac{\partial }{\partial t}\hat{q}\psi$. Carrying out, then, the substitutions, and integrating by parts, we have
\begin{equation}\label{eq:nick}
\begin{aligned}
    \CS_0[\psi]&=\int_0^{\frac{2\pi}{\omega_n}} \int \hat{p}\psi\frac{\partial}{\partial t}\hat{q}\psi dq dt\\
    &=\int_0^{\frac{2\pi}{\omega_n}} \int -\I  K\frac{\partial\psi}{\partial q}\frac{\partial}{\partial t}(q\psi) dq dt\\
    &=\int_0^{\frac{2\pi}{\omega_n}} \int -\I  K\frac{\partial\psi_n}{\partial q}q (-\I  \alpha \omega_n) \psi_n dq dt\\
    &= -2\pi K\alpha \int \frac{\partial\psi_n}{\partial q}q \psi_n dq\\
    &=-2\pi K \alpha \lim_{V\to \mathbb{R}^3}\int_{\partial V} \psi^2 q d\vec{S} + 2\pi K \alpha \int\psi_n^2 dq + 2\pi K \alpha \int \psi_n q \frac{\partial\psi_n}{\partial q} dq\\
    &=2\pi K \alpha - \CS_0[\psi],
\end{aligned}
\end{equation}
where we have used, as Schrödinger did in \eqref{SchrMinII}, that the boundary integral vanishes, namely, \begin{equation}\lim_{V\to \mathbb{R}^3}\int_{\partial V} \psi^2 q d\vec{S}=0,\end{equation} and the normalization of the wave function. Hence,
\begin{align}
   2 \CS_0[\psi]&=2\pi K\alpha=\alpha h.
\end{align}

In modern notation, \eqref{eq:nick} reads
\begin{align}
2\CS_0[\psi]&=\int_0^{\frac{2\pi}{\omega_n}}\ip{\psi}{\hat{p}\frac{\partial}{\partial t}\hat{q}\psi} +\overline{\ip{\psi}{\hat{p}\frac{\partial}{\partial t} \hat{q}\psi}}dt\\
&=\int_0^{\frac{2\pi}{\omega_n}}\ip{\psi}{\hat{p}\frac{\partial}{\partial t}\hat{q}\psi}   +\ip{\psi}{\left(\hat{p}\frac{\partial}{\partial t}\hat{q}\right)^\ast\psi} dt\\
&=\int_0^{\frac{2\pi}{\omega_n}}\ip{\psi}{\left(\hat{p}\hat{q}-\hat{q}\hat{p}\right)\frac{\partial}{\partial t}\psi} dt.
\end{align}

Now if we rewrite the foregoing expression for $\psi=\psi(x,t)=\psi_n(x)\e^{-\I  \alpha \omega_n t}$\footnote{Using \eqref{eq:PCFTranfII}, one sees that the decomposition of $\psi$ in such a product corresponds to the additive separation of variables $\tilde{S}(x,t)=K\log \psi(x,t)=K\log\psi_n(x)-\I  K\alpha\omega_n t=S(x)-\I  E_nt$, exactly as assumed by Schrödinger.}, we have
\begin{align}
    2 \CS_0[\psi]&=\ip{\psi}{[\hat{p},\hat{q}]\psi}\int_0^{\frac{2\pi}{\omega_n}}(\I\omega_n \alpha)dt\\
    &=\ip{\psi}{[\hat{p},\hat{q}]\psi}(2\pi \I \alpha),
\end{align}
from which it follows that
\begin{equation}\label{eq:Scrodinger-oldquantum}
    2\CS_0[\psi]=\alpha h \Leftrightarrow [p,q]=-\I  \hbar.
\end{equation}

This proves that Schödinger's procedure, applied to the reduced action $\CS_0$, \textit{recovers the Bohr-Sommerfeld rule with the condition that the commutation relations are satisfied}. Since Matrix Mechanics deduced the commutation relations precisely from $\CS_0$ by assuming the Bohr-Sommerfeld rule (see \S 2.1)\footnote{This is manifest in the fact that Born's substitution rule \eqref{quantumdJ} for $\Phi=J$ implies $\alpha h=J(n+\alpha)-J(n)$.}, we can state the conclusion: \\

\noindent \textit{the reason the canonical commutation relations are a common feature of Matrix and Wave Mechanics is that these theories are dual from the point of view of the action}. Each theory is constructed upon a different term of the same action and hence they are---and this is what we mean by ``dual''---\textit{two sides of the same thing}. \\

This is the main result of this section. We summarize it in the following
\begin{scholium} \label{sch:duality} Let the action $\CS[\psi]=\CS_0[\psi]-\CJ[\psi]$ be as above. The following two conditions are equivalent:
\begin{enumerate}[(i)]
    \item \label{item:wavemech} $(\delta \CS[\psi])_{\delta \CS_0[\psi]=0} =0$.
    \item \label{item:matrixmech} For every stationary state $\psi$, i.e. $\delta \CS[\psi]=0$, there exists $n\in \mathbb{N}$ such that $2\CS_0=nh$.
\end{enumerate}
Wave Mechanics (item (\ref{item:wavemech})) and Matrix Mechanics (item (\ref{item:matrixmech})) are therefore ``dual'' theories because they were constructed upon equivalent conditions.
\end{scholium}

We now want to use this result as basis for a critical analysis of the mathematical foundations of orthodox quantum mechanics.

\section{The Dynamical Duality of ``Unified'' Quantum Mechanics}

John von Neumann's \textit{The Mathematical Foundations of Quantum Mechanics} is arguably quantum-mechanics-as-we-know-it. We ask the reader to consider, in this connection, the following facts.\\ 

1) von Neumann presented his mathematical treatise as a ``unification'' of the ``equivalent'' Matrix and Wave Mechanics; 

2) It is with von Neumann's treatise that the most intensely investigated problem in the foundations of quantum mechanics, namely the ``measurement problem,'' or ``problem of outcomes'' \parencite{Maudlin1995}, is traditionally associated in the literature.\\ 

The main purpose of this section is to investigate, first, the link between these two facts, and then discuss the connection with our results.

\subsection{The Equivalence of the Two Theories: Hilbert Space?}

We want to begin by making explicit the specific tactics we have adopted. First, we shall indicate just what von Neumann's mathematical task, as he sees it, is---and for this, a couple of quotations from \parencite[\S I]{vonNeumann1932} will have to suffice; second, we shall unpack, since von Neumann did not, the mathematical burden of the task. Our own task will then be straightforward: Verify critically, in logical and mathematical terms, just how the burden was met in the body of the book. The logical outcome of this will be a reassessment of the mathematical foundations of quantum mechanics.\\ 

Consider the following statement from \S I.1, entitled ``The Origin of the Transformation Theory'':

\begin{quote}
A procedure initiated by Heisenberg [i.e., Matrix Mechanics] was developed by Born, Heisenberg, Jordan, and a little later by Dirac, into a new system of quantum theory, the first complete system of quantum theory which physics has possessed. A little later Schrödinger developed the ``wave mechanics'' from an entirely different starting point. This accomplished the same ends, and soon proved to be equivalent to the Heisenberg, Born, Jordan and Dirac system (at least in a mathematical sense, cf. 3 \& 4 below [here a footnote refers the reader to Schrödinger's equivalence paper]). On the basis of the Born statistical interpretation of the quantum theoretical description of nature, it was possible for Dirac and Jordan to join the two theories into one, the ``transformation theory,'' in which they make possible a grasp of physical problems which is especially simple mathematically. \parencite[6]{vonNeumann1932}
\end{quote}

Now if one is interested, as we are, in pinning down the facts involved in engineering a ``unified'' theory of quantum processes based on the equivalence of Matrix and Wave Mechanics, one has to look beneath von Neumann's remark that it was Born's statistical interpretation that allowed Transformation Theory ``to join the two theories into one.'' For Born introduced, as is well-known, the statistical interpretation of the wave-function in his papers on collisions and on a paper about the adiabatic theorem, all of which were published in June-October 1926 (see \parencite{guidostatistics2022} for the interesting details)---Schrödinger, let us recall, published his equivalence paper in March 1926: thus \textit{before} Born's papers on collisions. Now consider, with this in mind, the fact that therein Schrödinger showed that the isomorphism $f\mapsto (f_z)_{z\in Z}$ defined by $f(x)=\sum_{z\in Z} f_z \psi_z(x)$ for all $x\in \Omega$, where $\psi_z$ are an orthonormal basis of solutions to Schrödinger's equation, links the matrix-elements of Matrix Mechanics---which, with respect to their indices, ``taken singly, specify the \textit{states}, and in pairs, specify the \textit{transitions}'' \parencite[339]{Heisenberg&Born&Jordan1925}---, to the Fourier components of the wave functions of Wave Mechanics. We already remarked in \S2.3 upon the fact that Schrödinger made explicit use of this isomorphism to calculate the intensities of the Stark effect patterns in \parencite[\S4]{schrproper3}---a paper also published before Born's first paper on collisions, in June 1926. In ``The Exchange of Energy According to Wave Mechanics,'' Schrödinger's last paper on Wave Mechanics, published in June 1927, Schrödinger stated the obvious: His rule to compute the intensities---which, let us repeat expressly once more, he had learned from Matrix Mechanics---, and Born's statistical interpretation of the wave function, are mathematically equivalent.

Formally speaking, therefore, the Born rule was actually anticipated by Schrödinger---for the rule, after all, \textit{as Schrödinger himself described it}, is ``nothing but the 'translation' into the language of [wave mechanics] of very well-known considerations which Bohr brought forward in connection with calculation of line intensities by means of the principle of correspondence'' \parencite[91]{schrproper3}; ``the method of calculation [of the intensities] \textit{corresponds exactly} to the \textit{assumptions} of Born, Jordan, and Heisenberg'' \parencite[92, italics ours]{schrproper3}.\footnote{\label{foot:schrodingerstatistical}Note that these quotations apparently indicate, apparently \textit{entail,} that Schrödinger could not deny the statistical interpretation without self-contradiction, insofar as Bohr's correspondence principle, and the assumptions of Born, Jordan, and Heisenberg, cannot lead to a different interpretation (see \S1.1). That the contradiction is however only apparent is evinced by our remarks on footnote \footref{foot:11}, above.} The rule, that is, is a straightforward mathematical consequence of Schrödinger's equivalence proof.\footnote{See the definition of the correspondence principle on \cpageref{correspondenceprincipleIni,correspondenceprincipleEnd}. This conclusion is in disagreement with some of the details of \parencite{guidostatistics2022}---but the facts described as ``odd'' by Abraham Pais in the Postscript to his ``Max Born's Statistical Interpretation of Quantum Mechanics'' are in full harmony with it:
\begin{quote}
 Jorgen Kalckar from Copenhagen wrote to me about his recollections of discussions with Bohr on this issue. ``Bohr said that as soon as Schrödinger had demonstrated the equivalence between his wave mechanics and Heisenberg's matrix mechanics, the 'interpretation' of the wave function was obvious.... For this reason, Born's paper was received without surprise in Copenhagen. 'We had never dreamt that it could be otherwise,' Bohr said.'' \parencite[1198]{PAISONBORN}   
\end{quote}
See also footnotes \footref{foot:11},\footref{foot:equivalencepaper}, \footref{foot:schrodingerstatistical}.} 

This is the first main conclusion of this section. It shows that when von Neumann remarked that it was Born's statistical interpretation that allowed ``joining the two theories into one,'' he was, in a sense, turning the tables. For the mathematical fact of the matter is that once Schrödinger ``joined the two theories into one,'' from a formal point of view, the Born rule was unavoidable.\footnote{\label{foot:statisticalint}The key, most important theoretical fact involved in Born's work on the statistical interpretation and in the formulation of the so-called Born rule is his proof, in ``On the Adiabatic Principle of Quantum Mechanics,'' of the Adiabatic Theorem. It is an extremely important, but half-forgotten, fact that Ehrenfest's adiabatic principle had been used by Bohr since as early as 1916 to fix the stationary states of quantum systems and justify the \textit{a priori} status and invariance of the transition probabilities; see \parencite{EhrenfestonBohr1923} for a discussion of the general facts, and \parencite[\S5.2]{duncan2019constructing} for a good description of Ehrenfest's work. The methods of calculation available until 1926 were however restricted to periodic and conditionally-periodic systems; what Born achieved in doing, by taking advantage of Schrödinger's equivalence proof, and using the wave equation to determine the stationary states and estimate the transition probabilities for collision processes, is to extend the well-known procedure for non-periodic systems not only quantitatively (this Schrödinger had done himself, as we have seen) but also qualitatively, with his proof of the adiabatic theorem. The adiabatic theorem proved by Born was indeed a tremendous achievement, and a hallmark in the history of quantum mechanics, inasmuch as it \textit{completed} the statistical program Bohr had started in 1916---hence von Neumann's remark. We shall discuss the mathematical and 
physical connection between Ehrenfest's adiabatic principle, and the dual action, in a separate paper. }

Good. To proceed, then, with von Neumann's argument: in \S I.3, called ``The Equivalence of the Two Theories: The Transformation Theory,'' von Neumann remarks that against Dirac and Jordan's Transformation Theory stands the determinative requirement of mathematical rigor:
\begin{quote}
We do not desire to follow any further here this train of thought which was shaped by Dirac and Jordan into a unified theory of the quantum processes. The “improper” functions (such as $\delta(x)$, $\delta'(x)$) play a decisive role in this development—--they lie beyond the scope of mathematical methods generally used, and we desire to describe quantum mechanics with the help of these latter methods. \textit{We therefore pass over to the other (Schrödinger) method of unification of the two theories.} \parencite[20, italics ours]{vonNeumann1932}
\end{quote}

In \S I.4, a section entitled ``The Equivalence of the Two Theories: Hilbert Space,'' von Neumann then argues that the reason the Dirac-Jordan Transformation Theory runs into trouble is that it tries to relate the ``discrete'' space $Z=\{1,2,3,\ldots\}$ of index values, the arena of Matrix Mechanics, with the continuous state space $\Omega\subset \mathbb{R}^d$, the arena of Wave Mechanics, but this is impossible. The spaces are completely unrelated. \textit{But}, he says, the \textit{functions} in these spaces---the space $F_Z$ of complex sequences $(x_z)_{z\in Z}$ and the space $F_\Omega$ of wave-functions $\Phi:\Omega \to \mathbb{C}$---are isomorphic.\footnote{The space-state $F_Z$ in Matrix Mechanics corresponds to set of square summable sequences $\ell^2(Z,\mathbb{C})$; meanwhile, in Wave Mechanics, the space-state corresponds to the set of complex square-integrable functions $L^2(\Omega,\mathbb{C},dx)$ on the measurable subset $\Omega \subset \mathbb{R}^d$. It turns out that $L^2(\Omega,\mathbb{C},dx)$ is a separable Hilbert space with the usual inner product $\ip{f}{g}=\int_{\Omega} \overline{f(x)}{g(x)}dx$ and all separable Hilbert spaces are unitarily isomorphic by considering two orthonormal basis $\{e_n\}_{n\in \mathbb{N}} \subset \mathcal{H}_1$ and $\{f_n\}_{n\in \mathbb{N}} \subset \mathcal{H}_2$, and the linear map such that $e_n\mapsto f_n$ for all $n\in \mathbb{N}$.} It is the fact that Schrödinger related matrices to functions in his equivalence paper that is behind the italicized sentence in the passage quoted above: in \parencite[22, footnote 35]{vonNeumann1932} von Neumann remarks that Schrödinger's equivalence proof corresponds to the part of the theorem proved by Fischer and Riesz on the isomorphism of $F_Z$ and $F_\Omega$ which Hilbert proved in 1906. ``But Schrödinger didn't really establish the equivalence of Matrix and Wave Mechanic,'' the reader may now object. True---but the glitz of this mathematical fact should not blind us to the historical fact of the matter: that Schrödinger's contention that ``from the formal mathematical standpoint, one may speak of the \textit{identity} of the two theories'' \parencite[46]{schequivalence} was \textit{accepted}. For it is Schrödinger's contention, and the full Riesz–Fischer theorem, that are von Neumann's starting points:

\begin{quote}
Since the systems $F_Z$ and $F_\Omega$ are isomorphic, \textit{and since the theories of quantum mechanics constructed on them are mathematically equivalent}, it is to be expected that a unified theory, independent of the accidents of the formal framework selected at the time, and exhibiting only the really essential elements of quantum mechanics, will then be achieved if we do this: Investigate the intrinsic properties (common to $F_Z$ and $F_\Omega$) of these systems of functions, and choose these properties as a starting point. \parencite[24, italics ours]{vonNeumann1932}
\end{quote}
von Neumann is therefore clearly taking the equivalence between Matrix and Wave Mechanics \textit{as a given}. The isomorphism between $F_Z$ and $F_\Omega$ is also given; only one must note, and think through the fact, \textit{that such givens are independent}. For kinematical equivalence---let us repeat it expressly once more---does not imply dynamical equivalence. 

The isomorphism between $F_Z$ and $F_\Omega$ is a relation between \textit{static} structures; the Hilbert space itself \textit{contains no information about dynamics} (``the dependence on $t$ is not to be considered in forming the Hilbert space'' \parencite[128]{vonNeumann1932}). Since it must thus be acknowledged that the equivalence of Matrix and Wave Mechanics is not encoded in the abstract structure of the Hilbert space, to \textit{realize}, then, the ``expectation'' that a ``unified theory [...] exhibiting only the really essential elements of quantum mechanics'' should be erected over Hilbert space, von Neumann must present \textit{independent arguments} to prove that the equivalence between Matrix and Wave Mechanics---their complete dynamical parallelism from the mathematical viewpoint---does follow of necessity from the Hilbert space. \\  

We will follow this conclusion as a heuristic principle in our analysis of von Neumann's work. It is in connection with it that we now want to recall that von Neumann presented a proof of the Stone-von Neumann theorem in \S II.9 of his treatise; this theorem, which is too well-known to warrant description, is referred to in the literature as ``the mathematical reason behind the equivalence of the Heisenberg quantum mechanics and the Schrödinger wave mechanics'' \parencite[333]{emch1984mathematical}. Let us assess this statement.

\begin{enumerate}
    \item The Stone-von Neumann theorem clearly doesn't shed any light whatsoever on how it can \textit{be} that Heisenberg's and Schrödinger's apparently orthogonal initial assumptions turn out to be different aspects of the same thing---unlike our \eqref{sch:duality};
    
    \item The Stone-von Neumann theorem is a result about the unitary equivalence of irreducible representations of the canonical commutation relations, which entails unitarily equivalent \textit{algebras}, which entails a unitary map linking the Hamiltonian $H(p,q)$ of Wave Mechanics to Matrix Mechanics and conversely. But consider the facts described in \S2.1-3---it will immediately appear that unitarily equivalent Hamiltonians does not establish dynamical equivalence between Matrix and Wave Mechanics. For Matrix Mechanics posited, besides a unitary evolution for the Hamiltonian matrix $H$, that individual transition processes happen in discontinuous jumps to stationary states. The Stone-von Neumann theorem, therefore, is not---it cannot be--- ``the mathematical reason behind the equivalence of the Heisenberg quantum mechanics and the Schrödinger wave mechanics.'' There is more to the dynamics of quantum mechanics than the Hamiltonian.
\end{enumerate}

This establishes that the Stone-von Neumann theorem alone \textit{is not sufficient to establish the equivalence of Matrix and Wave Mechanics}. This is the second main conclusion of this section. It is with it in the back of our thinking that we now with to consider critically the following facts.

In \parencite[\S III]{vonNeumann1932} von Neumann proposed that the evolution of a quantum system occurs in two stages which he called ``process 1'' and ``process 2'': a non-unitary, stochastic ``jump'' induced by measurements, and a unitary evolution described by Schrödinger's time-dependent equation of motion between measurements. This duality---neither encoded in the static Hilbert space structure, nor implied by the Stone-von Neumann theorem---must then be justified. We will now consider how von Neumann framed the problem, and stated his answer, in chapters V and VI respectively. 

We read in chapter V:
\begin{quote}
We must now analyze in greater detail these two types of [processes]—--their nature, and their relation to one another. First of all, it is noteworthy that [process] 2 admits of the possibility that H is time-dependent, \textit{so that one might expect that 2 would suffice to describe interventions caused by measurement}: indeed, a physical intervention can be nothing other than the temporary insertion of a certain energy coupling into the observed system; i.e., the introduction into H of a certain time dependency (prescribed by the observer). \textit{Why then have we need—for measurements—of the special process 1?} The reason is this: In a measurement we cannot observe the system S by itself, but must rather investigate the system S + M in order to obtain (numerically) its interaction with the measuring apparatus M. The theory of measurement is a statement concerning S + M, and should describe how the state of S is related to certain properties of the state of M (namely, the positions of a certain pointer, since the observer reads these). Moreover, it is rather arbitrary whether one includes the observer in M, and replaces the relation between the S state and the pointer positions in M by relations between this state and chemical changes in his eye or even in his brain (i.e., to that which he has “seen” or “perceived”). We shall investigate this more precisely in VI.1. In any case, the application of 2 is of importance only for S + M. Of course, we must show that this gives the same result for S as does the direct application of 1 to S. If this proves successful, then we will have achieved a unified way of looking at the physical world on a quantum mechanical basis. \parencite[230, italics ours]{vonNeumann1932}   
\end{quote}
The measurement problem, as von Neumann sees it, is therefore this:
\begin{description}
\item [MP Formulation $\alpha$.] Why does process 2 not suffice, in practice, to describe measurements, given that it could in principle (if $H$ is time-dependent)? Why do we need process 1 for measurements?   
\end{description}
Since the completeness of the wave-function, and the eigenvector-eigenvalue link, are implicitly presupposed, we find in MP Formulation $\alpha$ (henceforth, MP $\alpha$) the historical origin of the so-called problem of outcomes \parencite{Maudlin1995} and of the view von Neumann introduced process 1 to solve it. Now consider against this fact the further fact that when von Neumann returns to the question in VI.1, ``The Measurement Process: Formulation of the Problem,'' he puts the matter in different, more general, terms:
\begin{quote}
In the discussion so far we have treated the relation of quantum mechanics to the various causal and statistical methods of describing nature. In the course of this, we have found a peculiar dual nature of the quantum mechanical procedure which could not be satisfactorily explained. \parencite[269]{vonNeumann1932}
\end{quote}
Here, then, the measurement problem reads:
\begin{description}
    \item [MP Formulation $\beta$.] What is the explanation for the dual nature of quantum dynamics?
\end{description}

MP $\alpha$ thus makes explicit a fundamental assumption that MP $\beta$ does not, namely that it is theoretically conceivable that process 2 should hold good for all processes. We are therefore logically justified to distinguish between such formulations. Furthermore, the fact that MP $\beta$ is more general than MP $\alpha$ makes it plain that it is not the measurement problem itself, but the ``solution'' von Neumann gives to it, that is investigated ``more precisely'' in chapter VI than in chapter V. For the answer in chapter VI reads (excuse the long quote):
\begin{quote}
First, it is inherently correct that measurement or the related process of subjective perception is a new entity relative to the physical environment, and is not reducible to the latter. Indeed, subjective perception leads us into the intellectual inner life of the individual, which is extra-observational by its very nature, since it must be taken for granted by any conceivable observation or experiment. Nevertheless, it is a fundamental requirement of the scientific viewpoint—--the so-called \textit{principle of psycho-physical parallelism}—--that it must be possible so to describe the extra-physical process of subjective perception as if it were in the reality of the physical world; i.e., to assign to its parts equivalent physical processes in the objective environment, in ordinary space. [...] In a simple example, these concepts might be applied as follows: We wish to measure the temperature. If we want, we can proceed numerically by looking to the mercury column in a thermometer, and then say: “This is the temperature as measured by the thermometer.” But we can carry the process further, and from the properties of mercury (which can be explained in kinetic and molecular terms) we can calculate its heating, expansion, and the resultant length of the mercury column, and then say: “This length is seen by the observer.” Going still further, and taking the light source into consideration, we could find out the reflection of the light quanta on the opaque mercury column, and the path taken by the reflected light quanta into the eye of the observer, their refraction in the eye lens, and the formation of an image on the retina, and then we would say: “This image is registered by the retina of the observer.” And were our physiological knowledge greater than it is today, we could go still further, tracing the chemical reactions which produce the impression of this image on the retina, and in the optic nerve and in the brain, and then in the end say: “These chemical changes of his brain cells are perceived by the observer.” But in any case, no matter how far we proceed—--from the thermometer scale, to the mercury, to the retina, or into the brain—--at some point we must say: “And this is perceived by the observer.” That is, we are obliged always to divide the world into two parts, the one being the observed system, the other the observer. In the former, we can follow all physical processes (in principle at least) arbitrarily precisely. In the latter, this is meaningless. The boundary between the two is arbitrary to a very large extent. In particular, we saw in the four different possibilities considered in the preceding example that the “observer”—--in this sense—--need not be identified with the body of the actual observer: in one instance we included even the thermometer in it, while in another instance even the eyes and optic nerve were not included. That this boundary can be pushed arbitrarily far into the interior of the body of the actual observer is the content of the principle of psycho-physical parallelism. But this does not change the fact that in every account the boundary must be put somewhere if the principle is not to be rendered vacuous; i.e., if a comparison with experience is to be possible. \parencite[272-273, italics in the original]{vonNeumann1932} 
\end{quote}

To this he adds:
\begin{quote}
Quantum mechanics describes events which occur in observed portions of the world, so long as they do not interact with the observing portion, and does so by means of process 2 (V.1). But as soon such an interaction does occur—--i.e., a measurement is made--—the theory requires application of process 1. This duality is therefore fundamental to the theory. [He adds in a footnote: ``Bohr was the first to point out that the duality which is necessitated by quantum formalism, by the quantum mechanical description of nature, is fully justified by the physical nature of things, and that it may be connected with the principle of psycho-physical parallelism.''] Danger, however, lies in the fact that the principle of psycho-physical parallelism is violated so long as it is not shown that the boundary between the observed system and the observer can be displaced arbitrarily, in the sense given above. \parencite[273]{vonNeumann1932}
\end{quote}
    
The problem of showing that the boundary between the system and the outside can be shifted arbitrarily, called by J. Bub ``the consistency problem'' \parencite[\S2]{Bub2001}, is solved by von Neumann in VI.2. But the solution to the consistency problem does not imply that the principle of psycho-physical parallelism is the correct answer to the measurement problem (as defined in MP $\alpha$). And the consensus in the international community of scholars today is that indeed \textit{it is not}. 

The reason is evident. To say that ``the duality which is necessitated by the quantum formalism [...] is fully justified by the physical nature of things, and that it may be connected with the principle of psycho-physical parallelism'' is to provide a metaphysical answer to a dynamical question. And to give to a dynamical question anything else but a dynamical answer is a mistake of a particularly infelicitous kind. It is a category mistake. 

The sophisticated discussions of scholars about von Neumann's answer to the measurement problem need not, however, detain us.\footnote{But see Abner Shimony's masterful \parencite{Shimony1963}.} What \textit{must} detain us is that which scholars have universally taken for granted, namely von Neumann's formulation of the question. For even though MP $\alpha$ has caused a flood of literature, the assumption that process 2 could, in principle, suffice for measurements if $H$ is time-dependent has always passed over as innocent. This is surely a massive oversight, for we find reason to suspect that it may not be in von Neumann's own words; a fundamental claim von Neumann made earlier in the book is in flat contradiction with it. And two contradictories cannot, by the laws of classical logic, both be true simultaneously. 

In \S2.1 we quoted a passage from chapter I of von Neumann's book wherein he emphasized that “all occurrences of an atomic molecular order obey the discontinuous law of quanta.” He reminded the reader at the outset of the fact that ``the general opinion in theoretical physics had accepted the idea that the classical principle of continuity is merely simulated in Nature;'' he underscored that we live in a world “which \textit{in truth} is \textit{discontinuous} by its very nature.” Now this implies as a matter of course that it is the certain effect of “the insertion of an energy coupling into the observed [quantum] system”---i.e., a finite disturbance of the system from the outside, \textit{or a measurement}---that the system will undergo a jumpwise transition. The contradiction mentioned above is now entailed by the simple fact that the Schrodinger equation, which describes the unitary dynamics, is a differential equation. It is based on the property of continuity which von Neumann asserts Nature itself lacks. So how can it really be \textit{thinkable} that the \textit{continuous} process 2 could in principle suffice to describe the \textit{discontinuous} transitions, as von Neumann says in MP $\alpha$? Must we not necessarily maintain, “so that one might expect that 2 would suffice to describe interventions caused by measurement,” that 2, \textit{and the interventions caused by measurement}, are continuous?

We are logically obliged to say that we must. But let us press the matter further; von Neumann’s inconsistency on this point can be unproblematical if, and only if, he is implicitly admitting the possibility that “the insertion of an energy coupling into the observed system” can happen continuously. Let us assume that this is the case; does MP $\alpha$ make sense then? We now categorically assert: \textit{it does not}. For the insertion of a time-dependency into $H$ does \textit{not} change the fact that the Schrödinger equation \textit{cannot} describe, \textit{even in principle}, “the insertion of an energy coupling into the observed system,” \textit{even if the world is continuous all the way down}, as the following remarks will readily demonstrate.\\

It is at this point that we bring the results of \S2 to the table. Condition \eqref{item:wavemech}, which is equivalent to Schrödinger's equation, holds only for $\delta \CS_0[\Psi]=0$, which means: \textit{no change in quantum number}. We saw in \S 2.2 that Schrödinger derived the time-independent wave equation in (1926b) by starting from the time-independent Hamilton-Jacobi equation; a straightforward calculation shows, however, that it suffices to start from the time-\textit{de}pendent Hamilton-Jacobi equation, and then follow further the same procedure, to obtain the time-\textit{de}pendent Schrödinger equation for a Hamiltonian that depends explicitly on $t$. In both cases the term $\CS_0$, which accounts for the transitions, is bypassed; what happens if H depends on $t$ in process 2 is only that besides the wave-function $\psi(x,t)$, the energy associated to the quantum numbers $n_t$ will then depend on $t$. But the condition $\delta \CS_0[\Psi]=0$---the point cannot be overemphasized---\textit{still holds}. If one inserts a time-dependency into $H$, \textit{it is still the case} that the wave equation \textit{cannot} describe the transitions between any two given quantum numbers $n_t$ and $m_t$ at any given time.\footnote{This is not to deny that it is in principle possible, given an initial state $\psi$, a final state $\phi$, and a time instant $t_1$,  to tailor a time-dependent Hamiltonian $H_{\phi,\psi,t_1}(t)$ such that the solutions of the Schrödinger equation satisfy $\psi(x,t_1)=\phi(x)$. We deny only that this constitutes a counter-example that can be cited as proof against the correctness of our argumentation. For not only to construct such a $H_{\phi,\psi,t_1}$ requires knowing the time $t_1$ and the final state $\phi$ in advance. It is also transparent that the described evolution from $\psi$ to $\phi$ is not a \textit{bona fide} transition.} Even though it is true that process 2 admits of the possibility that $H$ is time-dependent, the mathematical fact that one \textit{cannot} expect that 2 would suffice to describe interventions caused by measurement is not changed one iota; von Neumann's statement to the contrary is plausible only so long one ignores the premises from which the time-dependent Schrödinger equation is deduced. 

To protect ourselves against any charge of historical inaccuracy, we must point out to the fact that Schrödinger himself did not find the time-dependent wave equation in this way. In Schrödinger's own derivation, however, it is only $\Psi$, not $H$, that depends on $t$; Schrödinger, to make the dependence of $\Psi$ on the time $t$ explicit, merely eliminated the E parameter in the time-independent equation, upon using the trivial fact that for $\Psi(x, t) = \Psi(x)\e^{\pm \nicefrac{2\pi \I Et}{h}}$, $\nicefrac{\partial \Psi}{\partial t} = \pm \nicefrac{2 \pi \I}{h}E \Psi$. That everything that was said in the above paragraph holds for his derivation too is evident without explication. But this---mind you---Schrödinger explicitly \textit{said it himself}; an important, yet half-forgotten, fact is that even though Schrödinger begins \parencite{schrproper4}, the paper wherein he proposed the time-dependent equation derived in the way just mentioned, by saying that

\begin{quote}\label{Schrodinger-externalforces}
    there arises an urgent need for the extension of the theory for \textit{non-conservative} systems, because \textit{it is only in that way that we can study the behavior of the system under the influence of prescribed external forces}, e.g. a light wave, or a strange atom flying past. \parencite[103, italics ours]{schrproper4},
\end{quote} 
he ends the paper with the remark: ``I have not succeeded in forming [the wave equation] for the non-conservative case'' \parencite[123]{schrproper4}.

Our contention has thus been proved. It is not right, it is not legitimate, to expect that 2 would suffice to describe the interventions caused by measurements even if one omits the demand, which von Neumann said is imposed on us by Nature itself, that external perturbations give rise to discontinuous transitions. 

This is the third main conclusion of this section. It follows from it that MP $\alpha$ is, in a word, spurious. For if process 2 can't describe the interventions caused by measurement \textit{even in principle}, then it must be recognized that it makes no sense to ask why it doesn't do it in practice.\footnote{But there is more to this than meets the eye---we shall consider the facts that reveal the kind of game von Neumann was playing in stating MP $\alpha$ in \S3.2.} \\

So logic will now demand that we get rid of MP $\alpha$ and focus on MP $\beta$. It is the exact same question Muller named ``the measurement explanation problem'' in his \parencite{muller2023measurement}: 
 \begin{quote}
 The two postulates of standard QM that mutually exclude and jointly exhaust the change of state over time (Dynamics and Projection Postulate) evoke the question: why \textit{two}, and why \textit{these two?} \parencite[25]{muller2023measurement}
 \end{quote}
 Let remind ourselves why von Neumann was forced to ask MP $\beta$ to begin with. The reason lies in the heuristic principle defined a few pages back: The Hilbert space formalism of quantum mechanics was designed to be a self-standing structure independent of the ``accidents'' of the formalisms of Matrix and Wave Mechanics; the Hilbert space structure, however, being itself static, does not carry within itself the dynamical content of Matrix and Wave Mechanics. The dynamics of unified quantum mechanics had thus to be introduced independently from the outside. But unification requires proving the equivalence of Matrix and Wave Mechanics \textit{in Hilbert space}; the Stone-von Neumann theorem alone, however, does not suffice, as we have seen, to attain equivalence. Thus von Neumann could not adequately justify, \textit{on the basis of his framework alone}, why quantum mechanics should have the \textit{dual} dynamics that it does.

But our Scholium \ref{sch:duality} can.

Mind you, reader: If historically, Heisenberg, Born, and Jordan learned how to describe the evolution of the state of the isolated system only after Schrödinger presented his wave equation, and Schrödinger, by his turn, learned how to calculate the transitions only after he noted that the Fourier coefficients of the wave equation are isomorphic to Heisenberg’s matrix coefficients, so that the intensities are given by computing their square (Born rule), this was, it appears, no contingent development. By the constraints of the Scholium \ref{sch:duality}, such a course of events could hardly have been different as a matter of pure mathematics. Each rival theory started independently from the other, as we saw in \S2.1-2, \textit{by considering a different component of the action}; each theory then \textit{learned} from the other \textit{the missing piece of dynamical information required for a fuller dynamical description.} (\ref{item:wavemech}), the condition upon which Wave Mechanics was based, tells us that the unitary evolution holds only in so far there is no change of quantum number; (\ref{item:matrixmech}), the condition upon which Matrix Mechanics was based, prescribes the eigenvector-eigenvalue link (the quantum condition on the dynamically allowed states) and the collapse postulate (the “empirically given” jumpwise transitions). The first condition is mathematically and conceptually tied to the second by the bounds of the canonical commutation relations, \textit{via} $\CJ$ and $\CS_0$; the mathematical meaning of the relationship is clear: \textit{both conditions must be satisfied simultaneously if they hold individually.} 

This answers MP $\beta$ unequivocally. This tells us that the dynamics of a quantum mechanics based on the equivalence of Matrix and Wave Mechanics must have indeed \textit{two} dynamical postulates, and it tells us that it must be \textit{these two}.

MP $\beta$ is a general question about dynamics, and an adequate, accurate answer should itself be grounded on dynamics. This is precisely what the Scholium \ref{sch:duality} yields. It shows that and how the duality necessitated by the quantum formalism is itself necessitated by the dual nature of the quantum action.

This is the central conclusion of this section. We should like to remark in connection to it that in other branches of physics it is generally taken for granted that the dynamics of a physical theory should follow from the action functional. What we have disclosed here is the hitherto ignored fact that orthodox quantum mechanics is no exception. If orthodox quantum mechanics, based as it is on the equivalence of Matrix and Wave Mechanics, has this ``peculiar dual dynamics,'' this is so because one has been assuming a peculiar dual action.\footnote{It is perfectly true that the action pertains to the observable $H$ alone, whereas the form of the collapse depends on the particular choice of measurement. But our point remains unscathed, for any observable must evolve, irrespective of the choice of measurement, according to the temporal evolution of the formalism, and that is determined in orthodox quantum mechanics by the dual action.}

\subsection{The Equivalence of the Two Theories: Hilbert Space}

When Muller remarked, in the passage quoted above, that the two dynamical postulates of orthodox quantum mechanics exclude one another, he was expressing a very popular view. It is noteworthy that this view is closely related to the also popular view that von Neumann posited process 1 in an arbitrary, \textit{ad hoc} manner, only to solve the measurement problem. This affinity deserves our attention. For, since the dual action implies that the first view is mistaken, insofar as, according to it, the two dynamical postulates do not exclude, but in actual fact require, one another, it would be perplexing if the second view were nevertheless true. We want to expose the fact that it is not. \textit{Regardless of the antipathy that we may have for von Neumann's solution to the measurement problem}, it is illegitimate to criticize it on the grounds that it was arbitrary. It wasn't. And that should be made very clear.

The discontinuous, temporally undetermined transitions are the very idea that set the quantum revolution in motion; ``the fundamental laws [of the new mechanics],'' wrote James Jeans in 1914, in a influential book that reports the conclusions of the 1911 Solvay Conference, ``must be based on discontinuities, and not on the ideas of continuity involved in the classical mechanics'' \parencite[7]{jeans1914report}. Less well-known, because not emphasized by historians of quantum mechanics, is the fact that Matrix Mechanics---``a true theory of the discontinuum,'' as its authors called it---is the direct mathematical outgrowth of this firm conviction: that not only matter, not only electric charge, have a discontinuous, quantum structure, in the form of atoms and molecules and electrons.\footnote{The granular structure of matter was experimentally established by J. Perrin in 1908. Electrons were discovered by J. J. Thompson in 1897.} \textit{So does their motion, so does mechanics}. Thus Niels Bohr, who knew only too well that the lawfulness of continuity characterized the scientific and philosophical tradition that preceded him,\footnote{“The \textit{Law of Continuity} states that nature leaves no gaps in the orderings which she follows” \parencite[307]{von1996leibniz}. The lawfulness of continuity was also central to Hegel, Kant, and their heirs; see, e.g., \parencite{rogers1909}.} emphasized the point in every paper and every lecture:
\begin{quote}
Notwithstanding the difficulties which hence are involved in the formulation of the quantum theory, it seems, as we shall see, that its \textit{essence} may be expressed in the so-called quantum postulate, which attributes to any atomic \textit{process} an \textit{essential discontinuity}, or rather individuality, completely foreign to the classical theories, and symbolised by Planck's quantum of action. \parencite[88, italics ours]{bohrcomolecture}
\end{quote}
Now it is quite obvious that a mechanics based on continuity cannot admit discontinuities. A continuum admits, after all, of repeated or successive \textit{division without limit}, and all changes of \textit{state}, all changes of \textit{motion}, happen, in the mechanics of Newton, Hamilton and Legendre, passing through an infinitely divisible series of intermediate states. But it must be realized and carefully appreciated that as a matter of pure logic, a mechanics based on discontinuities cannot, unlike its counterpart, have a single, but it must have a dual, character. For without the continuum, discontinuities cannot even be \textit{defined}. Without the continuum, the discontinuities would not be \textit{discernible}. Thus it follows, since the old quantum theory was characterized by the conviction that the new mechanics should be based on discontinuities---for Planck's constant implies, as we all know, precisely in the impossibility of dividing a quantum-theoretical magnitude \textit{into ever smaller parts}, it implies precisely in the existence of an ultimate \textit{limit}---, that one could expect, indeed, that not only one, but \textit{two}, kinds of processes, were involved in the description of quantum processes \textit{already in the old quantum theory}: discontinuous and continuous ones; Bohrian quantum jumps should, then, ``grow out,'' so to speak, if this is right, of continuous processes.  Now, if there is any truth to these remarks, then we should be able to easily find textual evidence that corroborates them in the literature from the old quantum theory.  We now want to show to the reader that that is exactly the case. 

``The theory of [continuous] adiabatic transformations [for the quantum theory],'' wrote Ehrenfest in 1923,
\begin{quote}
was extensively clarified and deepened by Bohr's great work of 1918 [\parencite{Bohr1918}] and by the work he recently published \parencite{Bohr1923} on the basic postulates of quantum theory. [However,] as early as 1913 - in Part III of his epoch-making work ``On the constitution of atoms and molecules'' - Bohr had already used an adiabatic transformation: He thought a hydrogen molecule was formed by the gradual moving together of two neutral atoms and treated this process, which should take place very slowly against the orbit of the electrons, according to classical mechanics (analog: H + He, He + He). [...] Initially, Bohr preferred the term ``principle of mechanical transformability'' to the shorter one: ``adiabatic principle'' [...] [because] this designation emphasizes the fact that, according to this principle, \textit{quantum systems react quasi-classically under certain circumstances}, i.e., \textit{as if they obeyed classical mechanics}, if they are subjected to sufficiently smooth external influences---such as the ``adiabatic'' ones;\footnote{``If we consider a given conditionally periodic system which can be transformed in a continuous way into a system for which every orbit is periodic,'' says Bohr in the 1923 paper cited by Ehrenfest,
\begin{quote}
and for which every state satisfying \begin{equation}\label{eqquote1}\big[S_0=(x_1n_1+\cdots x_nx_n)h=nh\big]\end{equation} will also satisfy \begin{equation}\label{eqquote2}\big[(S_0)_k=n_kh \quad (k=1,\ldots, s)]\end{equation} for a suitable choice of coordinates, it is clear in the first place that it is possible to pass in a \textit{mechanical} [i.e. continuous] way through a continuous series of stationary states from a state corresponding to a given set of values of the $n$'s in \eqref{eqquote2} to any other such state for which $x_1n_1+\cdots+x_s n_s$ possesses the same value. If, moreover, there exists a second periodic system of the same character to which the first periodic system can be transformed continuously, but for which the set of $x$'s is different, it will be possible in general by a suitable cyclic transformation to pass in a mechanical [i.e. continuous] way between any two stationary states of the given conditionally periodic system satisfying \eqref{eqquote2}. \parencite[90, italics ours]{Bohr1923}
\end{quote}} but that, on the other hand, they \textit{may}, and (in general) \textit{should}, \textit{show their quantum claws}, as soon as the influence is no longer sufficiently patient and gentle.\footnote{In Bohr's own words:
\begin{quote} 
That mechanics, however, \textit{cannot generally be applied directly to determine the motion of a periodic system under influence of an increasing external field, is just what we should expect according to the singular position of degenerate systems as regards mechanical transformations}. [...] The process which takes place during the increase of the field will thus be analogous to that which takes place if an oscillating particle is subject to the effect of external forces which change considerably during a period. \textit{Just as the latter process generally will give rise to emission or absorption of radiation and cannot be described by means of ordinary mechanics, we must expect that the motion of a periodic system of several degrees of freedom under the establishment of the external field cannot be determined by ordinary mechanics}, \textit{but that the field will give rise to effects of the same kind as those which occur during a transition between two stationary states accompanied by emission or absorption of radiation}. Consequently, we shall expect that, during the establishment of the field, the system will in general \textit{adjust itself in some unmechanical way until a stationary state is reached} in which the frequency (or frequencies) of the above-mentioned slow variation of the orbit has a relation to the additional energy of the system due to the presence of the external field [...]. \parencite[89, italics ours]{Bohr1923}
\end{quote}} And Bohr shows us that this latter case exists as soon as a change in the motion conditions---however slow---leads from a ``degeneration'' with a certain ``degree of periodicity'' $u$ to a lower degeneration with a higher degree of periodicity $u'$ [...] \parencite[548-549]{Ehrenfest1923}
\end{quote}
The details of the theory of adiabatic transformations, and its role in the development of quantum mechanics, are beyond the scope of this paper. We shall look into it, and discuss its connection with the time-dependent Schrödinger equation and with the properties of the dual action, in a separate article.\footnote{See footnote \footref{foot:statisticalint}.} Here we want only to highlight the half-forgotten, but extremely important, fact that up to now the historical scholarship on the development of quantum mechanics has failed to bring out with full clarity: the fact that, \textit{before} the rise of Matrix and Wave Mechanics, since, indeed, \textit{at least as early as 1918}, and thus since at least as early as Bohr first proposed the statistical interpretation of quantum states \parencite{Bohr1918}, it was supposed, perhaps even \textit{expected}, that the new, quantum mechanics would involve to kinds of processes: a continuous, deterministic one, for small perturbations---``as if it obeyed the [deterministic] laws of classical mechanics''---and, for perturbations that ``are not sufficiently patient and gentle,'' discontinuous jumps.\footnote{``Bohr considered Schrödinger's investigations to be very important,'' wrote Heisenberg, reminiscing, decades later, about the effect of Schrödinger's equivalence papers in the discussions about the formalism of quantum mechanics in Copenhagen,
\begin{quote}
[for] Bohr realized at once that it was \textit{here} that we would find the solution to those fundamental problems with which he had struggled incessantly since 1913 [!], and in the light of the newly won knowledge, he concentrated all his thought \textit{on a critical test of those arguments which had led him to ideas such as stationary states and quantum transitions}. . . . I myself was not really willing to concede Schrödinger's theory a part in the interpretation of quantum theory. I considered it rather an extremely useful tool for solving the mathematical problems of quantum mechanics, but not more. Bohr, on the other hand, seemed inclined to place the wave-particle dualism among the basic assumptions of the theory. \parencite[101, italics ours]{Heisenberg1967}
\end{quote}
From the facts described above about the old quantum theory, it is clear that Bohr's position---unlike Heisenberg's---was, indeed, a direct consequence of previous developments.} The point we are leading up by way of all of this will now be obvious to the reader. These are the exact same two kinds of processes that von Neumann postulates and compares in his book.
\begin{quote}
We have then answered the question as to what happens in the measurement of a quantity $\mathcal{R}$ under the above assumptions for its operator $R$. [...] This discontinuous transition from $\Phi$ to one of the states $\psi_1, \psi_2,$ . . . (which are independent of $\Phi$ because $\Phi$ enters only into the respective probabilities $P_n = |(\Psi, \psi_n)|^2$, n = 1, 2, . . .) is certainly not of the type described by the time-dependent Schrodinger equation. This latter always results in a continuous change of $\Phi$, in which the final result is uniquely determined and is dependent on $\Phi$ (see the discussion in III.2). \parencite[140]{vonNeumann1932}
\end{quote}
In a footnote, von Neumann then reminds us: ``these jumps are related to the `quantum jumps' concept of the older Bohr theory'' \parencite[190, footnote 125]{vonNeumann1932}. 

Now von Neumann, with his characteristic rigor, does not fail to express the relation in exact mathematical terms.  Calculating Born probabilities, as we know, \textit{is} calculating the probabilities of measurement outcomes, and in chapter III, section 6, of his treatise, von Neumann obtains, \textit{using the Born rule,} the transition probabilities for the emission, absorption, and spontaneous emission of the radiation emitted by a quantum system that jumps from one stationary state to the other by describing the joint systems $S+L$ of the atom $S$ and the system consisting of the electromagnetic radiation $L$. The Hamiltonian of such a system is decomposed in three terms, $H=H_0+H_1+H_i$;  $H_0$ is the Hamiltonian of the (unperturbed) system $S$,  $H_1$ is the Hamiltonian of the (unperturbed) system $L$, and $H_i$ is the interaction between $S$ and the electromagnetic radiation. Following standard perturbation theory methods, von Neumann expands a general function 
\begin{equation}
f(\xi,u_1,\ldots,u_s)=\sum_{k=1}^\infty \sum_{\substack{m_0, m_1, \ldots=0\\ m_1+m_1+\cdots=s}}^\infty a_{km_0 m_1\ldots}\phi_k(\xi)\psi_{m_0,m_1\ldots}(u_1,\ldots,u_s)
\end{equation}
in the orthonormal basis $\{\phi_k\psi_{m_0,m_1\ldots} \ | \ 1\leq k\leq \infty, \ 0\leq m_j \text{ for all } j\in \mathbb{N},  \ \sum_{j=0}^\infty m_j=s\}$ of eigenfunctions of $H_0$, where $s$ refers to the number of ``light-quanta'' in the system $L$, initially supposed to be finite. Next he claims---and it is here that Born's rule appears explicitly---that 
\begin{quote}
    [f]or an arbitrary state $a_{k m_1 m_2 \ldots}$ of $S + L$, therefore, the configuration mentioned above (if it is measured: see the comments in III.3 on the non-degenerate pure discrete spectrum) has the probability [...] $|a_{k m_1 m_2 \ldots}|^2$.'' \parencite[185]{vonNeumann1932}
\end{quote}
\noindent He uses the time-dependent Schrödinger equation to obtain a differential equation for $a_{k m_1 m_2 \ldots}(t)$ on the system $S+L$ with Hamiltonian $H$, in which one can see that the terms $\frac{\partial}{\partial t}a_{k m_1 m_2 \ldots}$ depend indeed, in first-order approximation, only on $a_{k m_1 \ldots m_n\pm1 \ldots}$. And this inexorably leads, shows von Neumann, on short-time approximation, \textit{to the ``Bohr's frequency relation'' \eqref{Bohrenergycondition} for the emission or absorption of electromagnetic radiation}. And thus, says von Neumann, upon a measurement, and again, by means of Born's rule, ``the total probability that $S$ will be found in the $k$-th quantum orbit is 
\begin{equation}
    \theta_k=\sum_{m_1,m_2,\ldots} |a_{k m_1 \ldots m_n\pm1 \ldots}|^2 \quad\text{''},
\end{equation}
which expression leads, after calculation of the modulus of the coefficients $a_{k m_1 \ldots m_n\pm1 \ldots}$, to the probabilities of emission, absorption, and spontaneous emission of electromagnetic radiation (see \parencite[190-191]{vonNeumann1932} for the mathematical details).

Thus we can now see not only the historical, but also the \textit{theoretical}, relation between Bohr's quantum jumps and von Neumann's process 1: They are \textit{identical}, \textit{they are the same thing}, in the limit where one performs the measurement immediately after the interaction is turned on.

Now if anyone thinks that we are here defending von Neumann's process 1 as a solution to the measurement problem, let us emphasize, as strongly as we can, that this is not the case. It is historical truth that we are concerned with, and it must be admitted that the history of quantum mechanics is destroyed, and the comprehensibility of its conceptual development and structure in danger, whenever facts are no longer taken to be part and parcel of its past and present, and non-facts used to prove this or that claim or opinion. We hope that we are now impressed with the solid historical evidence that shows unequivocally that it is a misunderstanding, and the root of much misunderstanding, to discredit and dismiss von Neumann's work on the grounds that he introduced process 1 merely as an \textit{ad hoc} maneuver to solve the measurement problem. The truth is more complex, and indeed more interesting, than this. Not only von Neumann's process 1, but the dynamical duality he postulated in his formalism, have a prehistory that antedates his treatment of the measurement process.

In this way, we reach coherence with the facts described in \S3.1. For there we proved that it is mathematically illegitimate to expect that process 2 can in principle describe measurements, and this, in turn, implies admitting, \textit{contra} popular belief, that MP $\alpha$ is no genuine question. And the fact MP $\alpha$ is no genuine question is of course in perfect agreement with the fact that process 1 was not manufactured to solve it. 

But now a new problem confronts us. Now we must understand what was von Neumann's \textit{point}, then, when he formulated the apparently spurious, apparently vacuous, MP $\alpha$ and acted as though process 1 is the way out of the contradiction. For von Neumann must have had a point. It would be a grave injustice to him and his dignity as a mathematician to hastily infer that he must have been oblivious to the fact that the condition $\delta \CS_0[\Psi]=0$ holds for the time-dependent Schrödinger equation, that he, in formulating MP $\alpha$, unwittingly made a mistake. 

So let us think through the facts that will lead us to the solution of our puzzle. We know that Hilbert space isomorphism is static. And we saw that von Neumann's mathematical burden was then to present arguments that establish the equivalence between Matrix and Wave Mechanics \textit{in Hilbert space}; we know that Matrix and Wave Mechanics must have the same dynamics to be considered equivalent, and we saw, last but not least, that the Stone-von Neumann theorem \textit{alone} does not suffice to establish their dynamical equivalence. The Riesz–Fischer theorem, the Stone-von Neumann theorem, plus the collapse postulate---\textit{only then} equivalence between Matrix and Wave Mechanics can be attained in Hilbert space.  von Neumann's task as a mathematician, therefore, was to prove the necessity of the collapse postulate---but again, in Hilbert space.\footnote{See footnote \footref{foot:Bellimpossibility}.} To do so, he availed himself of one of a mathematician's favorite tactics, the \textit{reductio}. Having designed the linear structure to accommodate the discontinuous jumps,* he rhetorically asked what would happen if the jumps were eliminated. He then derived contradictions.

The Impossibility Proof is designed to establish the claim: ``quantum mechanics is in compelling logical contradiction with causality'' \parencite[213]{vonNeumann1932}. Since the quantum jumps are acausal, establishing acausality as an ineliminable feature of quantum mechanics automatically establishes that the discontinuous jumps are an ineliminable part of quantum mechanics. But look beneath this claim. The Impossibility Proof states this: It follows from the general validity of the expression $Exp(R) = \textbf{Tr}(UR)$ for the expected value of an observable $R$, given in terms of a density matrix $U$, that there are no completely dispersion-free states ensembles for quantum quantities, and since hidden variables are supposed to lead to dispersion-free ensembles, it follows that they are impossible. But note that von Neumann naturally assumed that the states associated to the hidden variables behave linearly. After all, otherwise, \textit{they would not fit into Hilbert space}. And note that this amounts to assuming that hidden variables are irrelevant for quantum mechanics to begin with, as pointed out by Grete Herrmann as early as 1933. For it follows from the argument---and this was von Neumann's whole point---that \textit{they would also undergo discontinuous transitions upon measurements}. The argument to make the acausal jumps appear inevitable is clearly circular, but note, reader, just how von Neumann camouflaged the circle. He camouflaged it in the way he stated the \textit{problem} the Impossibility Proof was designed to address. He contrasted the nature of the statistics of quantum mechanics with that of the kinetic theory of heat. ``Although we believe that after having specified we know the state of the system completely,'' he emphasized, ``nevertheless only statistical statements can be made concerning the values of the physical quantities involved'' \parencite[134]{vonNeumann1932}.

A little reflection will show that there's more to this comparison with the kinetic theory of heat than meets the eye. For the statistics and the dynamics are independent postulates in the kinetic theory of heat, and it goes without saying that determinism is a property not of the equations of state, \textit{but of the dynamics}. And so we will see, if we stop to think about it, that von Neumann's suggestion that it should be \textit{a priori} possible to implement causality in quantum mechanics depends logically, and therefore \textit\textit{necessarily}, on the \textit{a priori} possibility that the unitary time-dependent Schrödinger equation \textit{alone} exhausts the dynamics; the possibility of retaining determinism in his framework depends, in other words, on the \textit{a priori} possibility that process 2 can describe the transitions induced by measurement. And that is the exact same assumption von Neumann made, although in that case explicitly, in the statement of the measurement problem. 

Consider the claim that von Neumann's Insolubility Proof was designed to establish: ``the non-causal nature of the process 1 cannot be attributed to the incomplete knowledge of the observer'' \parencite[284]{vonNeumann1932}. The above considerations about the kinetic theory of heat apply in this case just as well; this conclusion of von Neumann's should follow from the dynamics of the theory, and from it \textit{alone}. But it turns out that by our analysis, \textit{that is precisely von Neumann's point}. For it is shown, in the Insolubility Proof, that the statistical behavior of quantum systems cannot be reproduced by an ignorance-interpretation mixture of the measurement apparatus M which interacts for some time with the measured system S according to process 2\footnote{See Harvey Brown's classic \parencite{Harvey1986}. See also the novel analysis given by Guido Bacciagaluppi in \parencite{Guido2013}.} And so it must be acknowledged---and this, let us repeat it expressly once more, is the essence of von Neumann's claim, the whole reason why he's going into this matter---that the statistic is a \textit{consequence} of the dynamics; process 1 is dynamical, full stop. von Neumann's Insolubility and Impossibility Proofs were therefore both designed to prove the mathematical necessity of process 1 as a dynamical postulate; the problem of causality, and the measurement problem, were the vehicles that gave to these proofs a logical foundation. For the reader knows how proofs by contradiction work. One first negates what one wants to establish and then shows that the consequences of this are not possible. The suggestion that process 2 can in principle describe measurements, or equivalently that process 1 can be dispensed with for S+M mixtures, is made to establish that the opposite is the case; it is calculated to make the general self-consistency point that ``the duality is therefore fundamental to the theory'' \parencite[273]{vonNeumann1932}. 

If this is right, then, \textit{and if one defines the measurement problem as the problem of outcomes}, as in \parencite{Maudlin1995}, which is nothing but an elaboration of MP $\alpha$,  it will now follow that the collapse postulate was not, as a matter of historical and mathematical fact, designed by von Neumann to solve the measurement problem. \textit{Instead the measurement problem was invented to justify the collapse postulate}. 

In this way we can understand the otherwise mysterious fact that not until the late 1950s---not, in point of fact, insofar as we could verify, until Hugh Everett's ``The Theory of the Universal Wave Function'' (1957)---one can find no reference in the literature to the contradiction defined by the problem of outcomes.\footnote{Einstein and Schrödinger always explicitly acknowledged the logical consistency of the formalism of quantum mechanics; so did David Bohm when presenting his hidden-variables interpretation (see the opening remarks of his seminal \parencite{Bohm1952}). Indeed, the first to rephrase MP $\alpha$ in terms of the now-familiar problem of outcomes, thence starting the trend of understanding ``interpretations'' of quantum mechanics as a way out of such contradiction, seems to have been Hugh Everett.} The shift in the foundational status of the temporally undetermined jumps, and the shift in the credibility of the so-called ``Copenhagen interpretation,''\footnote{The scare quotes are meant to represent the known fact that even though the Copenhagen interpretation is a \textit{fait accompli}  as far as textbooks and philosophical discussions are concerned, it is no unified point of view; see the details in \parencite{howard-thecopenhagenint}. We must, however, emphasize that we disagree with Howard in an important point: he claims that Niels Bohr did not subscribe to von Neumann's collapse postulate.} are historically connected. We can identify two components to this shift. 

The first is that today's scholars, unlike those from the pre-Bell period, have not been fed discontinuity with their mothers' milk; to these thinkers, the ``Copenhagen interpretation,'' which follows out the logic of this ``irrational'' concept (Bohr), sounds like pseudomystical, lawless nonsense. The second component is the ``formalism \textit{vs.} interpretation'' distinction myth.\footnote{See again footnote \footref{foot:statisticalint}.} For the myth had the effect of giving to von Neumann's edifice a façade of ``formalism''---as if physical theories could be invented without making, to start with, physical hypotheses; and since von Neumann pretended, for the reasons just explained, that it was \textit{a priori} possible that process 2 could describe the transitions induced by measurements, he thereby opened the possibility of inferring that it should be \textit{a priori} possible to get rid of the collapse postulate. 

In this way we can understand how it can be that the problem of outcomes seems, on a first and second glance, and out of historical context, a serious and legitimate formal difficulty, whereas in the context of von Neumann's construction it was a set up for \textit{reductio} proofs that, together with the Stone-von Neumann theorem, and the Riesz–Fischer theorem, assure, logically and mathematically, that Matrix and Wave Mechanics have indeed been, as they must be, \textit{ex hypothesi}, properly unified in Hilbert Space.

This closes our analysis of von Neumann's mathematical foundations of quantum mechanics. We want only to once again reiterate, to avoid misunderstandings, that our discussion must in no way, shape, or form be confounded with a defense of any version of the ``Copenhagen interpretation.'' Our aim here is only to set the facts straight, as we believe that we cannot overcome the \textit{fait accompli} of the Copenhagen heritage at their expense. And it is by all means a \textit{fact} that von Neumann's choice of dynamical postulates was not arbitrary. The dual dynamics is necessitated by the equivalence---kinematical and dynamical---of Matrix and Wave Mechanics, which he took as a given from the outset.\footnote{It is worth emphasizing here that other quantum theories---like Everett's, and de Broglie-Bohm's, for example---are not constructed on the basis of this assumption.} The further, deeper question is \textit{why} Matrix and Wave Mechanics are equivalent; why, in other words, the dual dynamics is fundamental to unified quantum mechanics. \textit{It is to this question that we claim to have found an answer.}  For whereas von Neumann only assumed that Matrix and Wave Mechanics are mathematically equivalent theories, and constructed unified quantum mechanics accordingly, we have been able to trace the matter a step further back, and found the dynamical duality entailed by the equivalence between Matrix and Wave Mechanics to have its origins in the dual nature of a certain action functional.

\section*{Acknowledgements}

We are grateful to Harvey Brown and Don Howard for helpful discussions in earlier stages of this investigation. We are also thankful to Nick Huggett, Guido Bacciagaluppi, and two anonymous referees, whose extraordinarily careful critical readings of earlier versions of this manuscript helped us correct mistakes and refine argumentation. The mistakes that remain are of course entirely our own.

\printbibliography
\end{document}